# Mechanism for collective cell alignment in *Myxococcus xanthus* bacteria

Rajesh Balagam[1], Oleg A. Igoshin[1]*

[1]Department of Bioengineering and Center for Theoretical Biological Physics, Rice University, Houston, Texas, United States of America

*Corresponding author
E-mail: igoshin@rice.edu

## Abstract

*Myxococcus xanthus* cells self-organize into aligned groups, clusters, at various stages of their lifecycle. Formation of these clusters is crucial for the complex dynamic multi-cellular behavior of these bacteria. However, the mechanism underlying the cell alignment and clustering is not fully understood. Motivated by studies of clustering in self-propelled rods, we hypothesized that *M. xanthus* cells can align and form clusters through pure mechanical interactions among cells and between cells and substrate. We test this hypothesis using an agent-based simulation framework in which each agent is based on the biophysical model of an individual *M. xanthus* cell. We show that model agents, under realistic cell flexibility values, can align and form cell clusters but only when periodic reversals of cell directions are suppressed. However, by extending our model to introduce the observed ability of cells to deposit and follow slime trails, we show that effective trail-following leads to clusters in reversing cells. Furthermore, we conclude that mechanical cell alignment combined with slime-trail-following is sufficient to explain the distinct clustering behaviors observed for wild-type and non-reversing *M. xanthus* mutants in recent experiments. Our results are robust to variation in model parameters, match the experimentally observed trends and can be applied to understand surface motility patterns of other bacterial species.

## Significance

Many bacterial species are capable of collectively moving and reorganizing themselves into a variety of multi-cellular structures. However, the mechanisms behind this self-organization behavior are not completely understood. The majority of previous studies focused on biochemical signaling among cells. However, mechanical interactions among cells can also play an important role in the self-organization process. In this work, we investigate the role of mechanical interactions in the formation of aligned cell groups (clusters) in *Myxococcus xanthus*, a model organism of bacterial self-organization. For this purpose, we developed a computational model that simulates mechanical interactions among a large number of model agents. The results from our model show that *M. xanthus* cells can form aligned cell clusters through mechanical interactions among cells and between cells and substrate. Furthermore, our model can reproduce the distinct clustering behavior of different *M. xanthus* motility mutants and is applicable for studying self-organization in other surface-motile bacteria.

## Introduction

*Myxococcus xanthus* is a model organism for studying self-organization behavior in bacteria [1]. These rod-shaped bacteria are known for their ability to collectively move on solid surfaces. Depending on environmental conditions, this collective movement allows cells to self-organize into a variety of dynamic multi-cellular patterns [2,3]. For instance, when nutrients are abundant, cells collectively swarm into surrounding spaces [1]. When cells come into direct contact with other bacteria that can serve as their prey, *M. xanthus* cells self-organize into ripples, i.e., bands of traveling high-cell-density waves [4-6]. Alternately, if nutrients are limited, cells initiate a multi-cellular development program resulting in their aggregation into 3-dimensional mounds called fruiting bodies [7,8].

Self-organization in *M. xanthus* requires coordination among cells and collective cell motility [1,5,6,9,10]. Despite decades of research, the mechanisms that allow for motility coordination in *M. xanthus* are not fully understood. In particular, the ability of cells to collectively move in the same direction is crucial to the observed multi-cellular behavior at various stages of their lifecycle [11-13]. Given that individual rod-shaped *M. xanthus* cells move along their long axis, coordination of cell direction in a group can be achieved by forming aligned cell clusters. Such clusters are observed in a variety of environmental conditions: low-density swarming [13], aligned high-cell-density bands in ripples [12] and long streams of aligned cells during the initial stages of aggregation [14,15]. However, the mechanisms responsible for this collective cell alignment are not completely clear.

Another important aspect of *M. xanthus* cell motility is the periodic reversal of its travel direction by switching the cell's polarity i.e., flipping the head and tail poles. Recent experiments indicate that the clustering behavior of *M. xanthus* cells is dramatically affected by variation in cell reversal frequency [16,17]. Starruß et al. [16] observed that, above a certain cell density, non-reversing *M. xanthus* mutants ( $A^+S^-Frz^-$ ) form large moving clusters, whereas reversing wild-type cells organize into an interconnected mesh-like structure. In a recent study, Thutupalli et al. [17] observed that starving wild-type *M. xanthus* cells increased their reversal frequency with time, which resulted in a change in their clustering behavior from aggregates (large clusters) to streams (elongated clusters). In addition, this study indicated that reversing and non-reversing cells differ in their dynamic behavior inside clusters. Reversing (wild-type) cells form stream-like clusters that appear stationary, and the cells move within the clusters. In contrast, non-reversing ( $\Delta frzE$ ) mutants form flock-like isolated clusters that move around, and the cells inside clusters appear to be moving with the same velocity as the clusters.

Therefore, our ability to explain cell alignment into clusters and variation of cell clustering behavior with changes in reversal frequency is essential for successful models of all self-organization phenomena. Several prior studies [16,18,19] attempted to understand the cell clustering process in *M. xanthus* using mathematical and computational approaches. Starruß et al. [16] developed a kinetic model, inspired from coagulation theory for colloidal particles, in which cell clusters' dynamics resulted from their fusion, splitting, and growth-decay processes. Using this model, they were able to explain the observed cluster size distribution for non-reversing cells. However, this model could not explain the cell clustering behavior for wild-type (reversing) cells. In another study, Harvey et al. [18] showed symmetry breaking between free cells (uniform gas phase) and nematically ordered cell clusters (dense phase) using a multi-phase continuum model. However, this model did not explicitly study the effects of changing reversal frequency on clustering, and the equations developed are limited to 1D and quasi-1D settings. Furthermore, both the models follow phenomenological approaches and do not provide a clear relationship between the model assumptions and individual cell behavior.

In this study, we overcome the limitations of previous approaches by connecting the individual cell behavior with collective cell motility through a biophysical agent-based model. Our overarching hypothesis is that cell clustering can be explained solely via mechanical interactions among cells and between cells and substrate. In other words, the observed patterns do not rely on biochemical signals such as chemotaxis. To test this hypothesis, we simulate interactions among a large number of cells through an agent-based simulation (ABS) framework. Using this framework, we first study the formation of aligned cell clusters in non-reversing *M. xanthus* cells and later extend our investigation to reversing cells. Furthermore, we investigate the effect of cell-substrate interactions, such as slime-trail-following, on the clustering patterns. The results of our simulation are compared with experimental data from the literature and can be applicable to other bacteria that display surface motility.

## Results

**Non-reversing flexible cells form clusters due to steric alignment**

First, we investigated whether mechanical interactions among *M. xanthus* cells would be sufficient to induce aligned cell cluster formation. This approach was motivated by our previous study [20], which demonstrated alignment in cell pairs as a result of head-to-side collision, and soft-condensed matter models for clustering in self-propelled rigid rod particles [21-24]. We hypothesized that successive collisions of cells with previously aligned cell clusters will result in the formation of even larger clusters. Thus, we simulated mechanical interactions among non-reversing cells, similar to self-propelled rod models, but with realistic cell flexibility values. For this step, we have used the bending stiffness value ($k_b$) for *M. xanthus* cells from our previous study [20], which reproduces realistic pair-wise cell collision behavior in model agents. Under these estimates of $k_b$, we studied clustering behavior of the model *M. xanthus* cells in our ABS framework at different cell densities ($\eta$, defined as the fractional area occupied by all cells in the simulation region).

To simulate mechanical interactions of cells moving on a 2D surface, we used our previously developed framework - briefly described below (see Methods for further details). In this framework, each agent consists of multiple segments, enabling a realistic mechanical model of a single *M. xanthus* cell. To this end, we use a connected string of nodes with linear and angular springs between nodes to simulate elastic behavior. Agents move forward through propulsive forces acting on the nodes tangential to the cell length (towards the next node). This is similar to the force generation through multiple motor protein complexes distributed along the cell length as observed by recent models of *M. xanthus* gliding motility [25-28]. Agents experience drag forces opposing their motion due to the surrounding fluid. Adhesive attachments to the underlying substrate at nodes resist lateral displacement of agents during collisions (the focal adhesion model of gliding motility [26]). At low densities, *M. xanthus* are known to move as a monolayer of cells. Therefore, collisions among agents are resolved by applying appropriate forces on nodes that keep agents from overlapping. Agents move over a 2D simulation space with periodic boundary conditions according to the net forces acting on their nodes. We introduce random noise in agent travel direction by altering the direction of propulsive force on the front node. We observe the agent behavior by solving Newton's equations of motion on nodes to obtain their position and velocity at each time step of the simulation. We use the Box2D [29] physics library to solve these equations of motion and efficiently handle the excluded-volume forces.

We start the simulation by initializing the cells one by one in the simulation region at random positions and with random orientations until the desired cell density is reached. While initializing, we accept only the cell configurations that do not result in cell overlap. As soon as the simulation begins, cells start moving and colliding with their neighboring cells and, as a result, align along their major axis [20]. This alignment is nematic [30]: aligned cells can move in the same or opposite directions depending on the initial orientation of cells. When aligned cells move in the opposite directions, they separate; however, when they move in the same direction, a small cluster of aligned cells is formed. These clusters grow in size as more cells join through collisions or due to merging with other cell clusters. Clusters shrink in size as peripheral cells leave the cluster due to random change in their travel direction (Movie S1, S2). We quantify the evolution of clusters through cluster size distribution (CSD, see Supp. Text). After approximately 180 min of simulation time, the CSD is stabilized (Fig S1), and we observe that cells in the simulation regions are distributed among clusters of different sizes, while few cells remain isolated.

Depending on the cell density ($\eta$), we observe a variation in the cluster size distribution and in the number of isolated cells. Cells form stable clusters (containing $>10^2$ cells) only for sufficiently high cell densities ($\eta \geq 0.16$) (Fig 1A-D), while cells largely remain isolated for lower densities ($\eta = 0.08$). We have quantified the effect of increasing cell density ($\eta$) on clustering behavior by measuring the mean cluster size $\langle m \rangle$ (refer to Supp. Text for details on the quantification procedures) at each cell density value. We observe that an increase in cell density results in an increase in mean cluster size (Fig 1E). We have quantified the alignment within the cell clusters using a mean cell orientation correlation, $C(r) = \langle \cos(2\Delta\theta_r) \rangle$, as a function of the neighbor cell distance $r$ (Fig 1F). Here, $\Delta\theta_r$ is the angle deviation between the orientations ($\theta$) of a pair of agents whose center nodes are separated by a distance $r$ (see Methods). We use $2\Delta\theta$ to ensure that correlation values in parallel and anti-parallel alignment configurations remain the same [31]. The orientation correlation results confirm that, in comparison with the initial distribution, clustering results in longer-distance orientation correlation for high cell densities. We observe that, immediately after the start of the simulation (1 min), cells exhibit very low correlation with their immediate neighbors ($r = 2-3\,\mu m$). However, after a long simulation time (180 min), we observe a large increase in cell orientation correlation with neighbor distances (except for $\eta = 0.08$, Fig 1F), indicating the formation of larger aligned clusters (refer to Fig S2 for the evolution of orientation correlation with time).

To test the robustness of our results, we have varied the cell flexibility ($k_b$) values over a wide range (0.1x – 10x) and studied the cell clustering behavior in our simulations. We observed that our model agents formed clusters except for the case of very high cell flexibility values (0.1x, $k_b = 10^{-18}\,N.m$) (Fig S3A-F). Furthermore, mean cluster sizes increased with increases in cell densities for all cell flexibility values (Fig S3G). Interestingly, increases in cell flexibility decreased the mean cluster sizes.

Thus, we observe that flexible agents can form aligned clusters through mechanical collisions for sufficiently high cell densities ($\eta \geq 0.16$), similar to self-propelled hard rods [19]. Furthermore, these cell clusters from our simulations are very similar to the isolated cell clusters experimentally observed for non-reversing *M. xanthus* ($frz^-$) cells [16,19].

**Periodic reversals destroy clustering**

Next, we investigated the effect of cell reversals on clustering behavior. We introduced periodic reversals of cell travel direction (reversal period = 8 min [32]) in our model agents. Similar to *M. xanthus* cells, each reversal results in a switch of the agent polarity i.e., flipping of the head and tail nodes. Surprisingly, with the addition of periodic cell reversals, cells failed to form large clusters even after a long simulation time (180 min) (Fig 2A, Movie S3). Furthermore, we observed that increases in cell density did not improve the mean cluster sizes significantly (Fig. 2B, black line). Even when we started with cells that initially formed clusters by simulating non-reversing cells first for 90 min and then turned on cell reversals, we observed the destruction of existing cell clusters within approximately 30 min (Fig S4). Thus, our simulation results indicate that steric alignment is not sufficient for formation of large aligned clusters in a population of periodically reversing agents. However, given that wild-type *M. xanthus* cells reverse their polarity but still form clusters, additional interactions must be included in our model to explain *M. xanthus* clustering behavior.

In our first attempt to correct this, we tested whether cohesive interactions among *M. xanthus* cells [33] can restore clustering. Studies on colloidal particles indicate that adhesion between particles can lead to their clustering [34]. *M. xanthus* cells secrete exopolysaccharide (EPS) proteins and fibrils on their surface, and these are observed to form a network with the surface fibrils of other cells that are in close contact, resulting in cell-cell cohesion [35,36]. These cohesive interactions can keep cells together and thus may lead to clustering in reversing *M. xanthus* cells. We investigated this mechanism by introducing lateral adhesion forces between neighboring agent nodes in our simulations (Refer to Methods). However, we observed that adhesive interactions between neighbor cells did not lead to significant cell clustering for reversing cells, even with high adhesion forces (Fig S5). Thus, lateral adhesions are not sufficient to stabilize the clusters of reversing cells.

To understand the rationale behind why cell reversals prevent the formation of large clusters, we examined the cell clustering dynamics in our simulations with and without cell reversals. For non-reversing cells, we observe that clusters grow in size due to collisions with new cells and that cells inside the clusters are unable to leave their cluster. At steady-state, cluster size is determined by a balance between the flux of peripheral cells leaving the cluster and new cells joining the cluster, similar to the kinetic theory developed in Ref. [19]. In contrast, for reversing cells, we observed that, even though mechanical collisions often lead to the transient formation of small clusters, these clusters fail to grow and stabilize. This occurs because, upon reversal, cells from the cluster interior move past the other cells in the opposite direction and leave the cluster. Furthermore, random changes in their travel direction prevent them from returning to their original clusters after another reversal. This also explains why adhesive cell interactions failed to result in the clustering of cells in our simulation. Lateral adhesive interactions do not stop cells from leaving the clusters after reversal and cannot influence the direction of cell movement once it leaves the existing cluster.

**Slime-trail-following by cells restored clustering for reversing cells**

Based on the results thus far, we conclude that an additional mechanism that could reduce random orientation changes in the cells could help overcome the destabilizing effects of reversals on clustering. A possible mechanism for this is suggested by the observation of slime-trail-following by *M. xanthus* cells. *M. xanthus* cells secrete slime, a polymeric gel, from their surface, and it is deposited on the underlying substrate as long trails during cell movement [37]. Furthermore, cells tend to follow their own trails after reversal, and, when in contact with slime trails deposited by others, cells can reorient and follow these [38]. Accordingly, we hypothesize that slime trails act as an orientation memory that reduces cells' ability to randomly change travel direction and assists in clustering for reversing cells.

We investigated the above mechanism of cell clustering based on slime-trail-following using our ABS framework. As the mechanistic basis of slime-trail-following by *M. xanthus* cells is not fully clear, we opt for a phenomenological model of slime-trail-following by reorienting part of the propulsive force on a cell's leading pole (head node) parallel to the slime trail it is crossing (Refer to Methods for more details). The results of these simulations indicate that the slime-trail-following mechanism restored clustering for reversing cells (Fig 2C, Movie S4). This is reflected by a significant increase in mean cluster sizes (green line in Fig 2B) for slime-trail-following cells compared to cells that do not follow slime trails (dashed line). Additionally, slime-trail-following also increased large-distance orientation correlations of cells, indicating the formation of aligned cell clusters (Fig 2D).

Notably, the cell clusters in our simulations for reversing cells with the slime-trail-following-mechanism resemble an interconnected mesh-like structure (Fig 2C). These clusters are distinct from the freely moving isolated cell clusters of non-reversing cells (Fig 1C). However, these interconnected cell clusters in our simulations are very similar to the interconnected mesh-like structure observed for wild-type (reversing) *M. xanthus* cells in experiments [16].

**Effective slime-trail-following and long slime trails required for clustering in reversing cells**

To investigate the robustness of clustering to the values of unknown parameters and to demonstrate key features of the model that are essential for clustering, we investigated effects of variation in the slime-trail-following ability of cells. For this, we perturbed the parameters that affect the slime-trail-following mechanism in our model: the slime effectiveness factor ($\varepsilon_s$), which controls the ability of a cell to follow a slime trail, and the slime trail length ($L_s$), which controls the memory effect of a cell path (refer to Methods for details). High $\varepsilon_s$ values decrease a cell's chance of escaping from the slime trail, whereas high $L_s$ values increase the chance of a cell to encounter slime trails from other cells. We have varied both parameters over a wide range in our simulations: $\varepsilon_s$ (0.1 to 1.0) and $L_s$ (0.16 to 11 μm).

For short slime trail length ($L_s = 0.16\,\mu m$) and a low slime effectiveness value ($\varepsilon_s = 0.1$), reversing cells show a dispersed cell pattern with minimal cell clustering (Fig 3A). This dispersed cell pattern is very similar to the situation for cells without slime-trail-following (Fig 2A). The underlying pattern of slime distribution in the inset shows minimal slime paths in the simulation, which do not effectively result in cells following others. Increasing the slime trail length to a higher value ($L_s = 11\,\mu m$) but keeping the slime effectiveness value low ($\varepsilon_s = 0.1$) did not improve cell clustering significantly (Fig 3B). Although cells are able to leave longer slime trails, creating an interconnected slime network (inset), the low slime effectiveness ($\varepsilon_s$) value allows cells to easily escape from the slime paths, and the slime-trail-following cannot effectively stabilize the formed clusters. In the same fashion, an increased slime effectiveness value ($\varepsilon_s = 1.0$) but a low slime trail length ($L_s = 0.16\,\mu m$) also did not result in significant cell clustering (Fig 3C). Here, even though cells are able to follow slime trails effectively, slime trails are not long enough for other cells to follow, and thus cells are more or less separated except for small cell clusters. However, with high slime effectiveness ($\varepsilon_s = 1.0$) and long slime trails ($L_s = 11\,\mu m$), cells are able to produce the normal cell clustering pattern for reversing cells (Fig 3D). Here, long slime trails allow for cells to follow other cells' slime trails, thus producing an interconnected slime network, and the high slime effectiveness factor prevents cells from escaping from slime paths and thereby results in a mesh-like clustering of cells. Thus, we observe that high slime-trail-following efficiency and sufficiently long slime trails allow for reversing cells to form cell clusters.

To further investigate the robustness of the slime-trail-following mechanism on agent clustering behavior, we have measured the mean cluster sizes via simulation for variations in slime effectiveness and slime trail length over a wide range of values ($\varepsilon_s = 0.1 - 1.0; L_s = 0.2 - 11\,\mu m$ - 64x change in slime production rate; see Methods for details). Our results indicate that except for very short slime trails ($L_s \leq 1\,\mu m$), increases in the slime effectiveness value increased the mean cluster sizes (Fig 3E). Similarly, increases in the slime trail length resulted in significant increases of mean cluster sizes except for very low slime effectiveness values (Fig 3F). Thus, reversing agents along with the slime-trail-following-mechanism can form clusters over a wide range of model parameters.

**Mechanical clustering model reproduces many features of observed *M. xanthus* cell behavior**

To further assess our clustering model, we decided to quantitatively compare our model predictions with the available experimental data on clustering behavior for both reversing and non-reversing strains of *M. xanthus*. To this end, we quantified the cell clustering behavior in our simulations by measuring the cluster size distribution, cell path maps, and cell visit frequency distribution from our simulations and compared our results with experiments reporting similar metrics [16,17].

First, we compared the cell cluster size distribution from our simulations with experiments of Starruß et al. [16]. For this, we performed simulations with the same cell density as in the experimental conditions for both reversing and non-reversing cells. We measured the cluster size distribution (CSD) from our simulations and plotted the probability, $p(m)$, of finding a cell in a cluster of size $m$ as a function of cluster size (solid lines in Fig 4A, B) and compared with the experimentally observed distribution (symbols). We observe that our simulation results qualitatively follow a similar trend to that of the experimental data. We chose model parameters (slime effectiveness, $\varepsilon_s$; slime trail length, $L_s$) to produce an approximate match. Global parameter optimization could further improve the agreement but was not performed. At small cell densities ($\eta = 0.08$), both reversing and non-reversing cells show a monotonically decreasing CSD with a large number of cells either being isolated or belonging to small clusters ( $m \sim 10-10^2$). However, no clusters larger than $10^2$ cells are observed. Nevertheless, with increases in cell density ($\eta$), non-reversing cells show a power-law distribution for CSD ($m^\beta, \beta = -0.90$ – closely matches with the result $\beta = -0.88$ from Starruß et al. [16]), and a significant number of cells now belong to large clusters ($m \sim 10^2 - 10^3$). In contrast, reversing cells show a decreasing CSD with increases in cluster size, and the largest clusters formed are limited to $< 400$ in size even at high cell densities.

Next, inspired by recent experimental studies indicating that wild-type (reversing) and ΔFrzE (non-reversing) *M. xanthus* mutants form distinct cell clusters that differ in their shape and dynamic behavior [17], we investigated these phenomena in our simulations. For this, we traced the cell paths over time and plotted the cell visit frequency of sites in the simulation region as a heat map for 2 consecutive hours after an initial transition period of 60 min (Fig 4C and D). We observed localized high-frequency visit areas and changing shapes of cell trace paths over time for non-reversing cells (Fig 4C), indicating the formation of large clusters that move all over the simulation region (Movie S5). In contrast, reversing cells organized into interconnected clusters that resemble a mesh-like structure, and the shape of the structure itself remained approximately the same over time (Fig 4D, Movie S6). Furthermore, the gap regions in the mesh structure (white areas) mostly remain free of cells or show very low visit frequency, indicating that reversing cells are confined within the cluster network (clearly seen for high-slime-trail-following-efficiency parameters, e.g., $L_s = 11 \mu m, \varepsilon_s = 1.0$; see Movie S4). Additionally, we have quantified the probability of cell visits, $p(N)$, as a function of visit frequency, $N$, in our simulations for both reversing and non-reversing agents(Fig 4E). We observe that simulations with reversing cells show a large fraction of sites with high visit frequencies ($N = 20 - 50$ visits for a 60-min interval) compared to non-reversing cells. Thus, reversing cells in the simulation region frequently visit specific sites, indicating stationary cluster structures. These results are qualitatively consistent with the observations of Thutupalli et al. [17] on the dynamic behavior of clusters.

## Discussion

Aligned cell clusters are crucial for formation of the multicellular structures observed during the *M. xanthus* lifecycle [12-15]. However, the mechanisms responsible for the cell alignment and clustering were not completely understood. Inspired by the studies of clustering in self-propelled hard-rods through mechanical collisions [21-24], we have developed an agent-based simulation framework to investigate mechanical collision-based cell clustering in *M. xanthus*. In this framework, each agent is based on a biophysical model of an individual *M. xanthus* cell that realistically mimics flexible cell motility behavior. The results from our simulations show that non-reversing flexible model agents can form clusters through mechanical collisions alone under realistic cell bending stiffness values of *M. xanthus* cells. However, the addition of periodic cell reversals eliminated the cell clusters in our simulations. Thus, we observe that mechanical collisions alone are insufficient for cell clustering of reversing cells. We hypothesized an additional mechanism of cell clustering based on slime-trail-following by *M. xanthus* cells. As expected, slime-trail-following by cells restored clustering for reversing cells. By varying the parameters in our model, we observe that effective slime-trail-following and long slime trails are required for cell clustering using the slime-trail-following mechanism. We quantified cell clustering behavior from our simulations and compared our results with experiments for both non-reversing and reversing cells. We observe that our simulation results qualitatively agree with experimental cell clustering behavior. Thus, our analysis shows that *M. xanthus* cells can form aligned clusters through mechanical collisions and slime-trail-following.

We believe that the following mechanism enables the reversing *M. xanthus* cells to form clusters through slime-trail-following (Fig 5A): a single *M. xanthus* cell leaves a slime trail while moving on a substrate and traces back its own trail while reversing and thus reinforces its own slime trail. When other cells cross this trail, they reorient and align with this slime trail and start following it. This results in a positive feedback mechanism where newly joined cells in the slime trail further reinforce the trail with their own slime, causing more cells to join the trail. Thus, more cells aligned with the original slime trail are recruited into the trail, resulting in a cluster of aligned cells. Within a cluster, cells maintain alignment with neighbor cells through mechanical interactions.

In the current study we limited cell densities ($\eta$) to 0.32 due to the limited availability of experimental data [16]. However, to extrapolate our conclusions, we have simulated the clustering behavior of cells for higher densities (up to $\eta = 0.60$). Results from these simulations indicate that cell alignment and clustering trough mechanical interactions also occur at these high densities (Fig S6). Interestingly we observe clustering of reversing cells at high cell densities even without slime-trail-following by cells (Fig S6B). These results suggest diminished role of slime trails in collective cell alignment at these conditions as the whole area covered by cells is likely to contain slime. However, we have opted not to investigate these conditions at greater depth due to limitations of our current 2D simulation framework and cluster quantification metrics for such conditions. At high densities cells in our simulations form large continuous clusters such that separating and characterizing individual clusters is practically impossible. Moreover at high cell densities real *M. xanthus* cells are capable of moving on top of one another resulting in a multi-layered biofilm whose dynamics are different from that of low cell density scenario. These effects would be explored in depth elsewhere.

Our simulations show that distinct clustering behaviors observed in *M. xanthus* mutant strains can be explained through mechanical interactions alone. Quantitative results from our simulations (CSD, cell visit frequency) follow the general trend as observed in experimental data [16,17]. Although our results do not exactly match with the experiments, this is understandable, as we were aiming to explain the observed cell clustering phenomena with a minimal interaction model. In our current model, we ignored many other interactions that exist among *M. xanthus* cells, e.g., the twitching of *M. xanthus* that uses type-IV pili to pull cells together. The addition of these processes along with further optimization of immeasurable parameters and choosing other model parameters from direct experimental observations (e.g., distribution of cell orientation changes, reversal time distribution) could further improve our current model but are beyond the scope of this study.

During development, *M. xanthus* cells exhibit circular aggregates, some of which later serve as initial fruiting body seed centers [14]. A recent study by Janulevicius et al. [39], using an agent-based-model similar to our current model, concluded that cells form circular aggregates when the end parts of leading and lagging cell pairs interact through short-range active forces that keep the distance between cell pairs constant. They reasoned that such active forces can come through type-IV pili at the leading end of a cell interacting with the other cell surface or through adhesive interactions between cell poles. However, in our current simulations, we occasionally observed such circular aggregates (Fig 5B) without using any active interactive forces between end-to-end cell pairs. Furthermore, in contrast to the predictions of [39], we observe that these aggregates do not rotate as rigid bodies as the agents inside the aggregate slide past one another (Movie S7). In our simulations, agents move with approximately the same speed, and, as a result, the angular velocity is higher for cells near the aggregate center. Thus, we argue that the circular aggregates observed in *M. xanthus* cells can be explained by slime-trail-following without active attractive forces between cells and propose that tracking cells in such aggregates can discriminate between the alternative models of their formation.

Cell clustering and the alignment of cells inside the clusters play a major role in *M. xanthus* physiology. *M. xanthus* are predatory bacteria that feed on other bacteria by secreting proteolytic enzymes into their surroundings. To maximize their predation, these cells form groups that move together. The alignment of cells inside these groups allows for a dense packing of cells per a given area, thereby increasing their predation efficiency. Furthermore, the variations in cell-clustering behavior observed by Thutupalli et al. [17] with concomitant changes in cell reversal frequency may enable starving cells to optimize their search for nutrients. During the initial phase of starvation, *M. xanthus* cells exhibit a low reversal frequency that allows them to form flock-like clusters that explore their surroundings for nutrients. Once nutrients are found, cells switch to a high reversal frequency, thus enabling cells to form stationary cluster structures that allow them to conduct optimal nutrient gathering.

Notably, cell clustering via slime following is observed in other bacterial systems. A recent study by Zhao et al. [40] showed that *P. aeruginosa* also uses a slime-trail-following mechanism to form initial cell clusters. Using cell-tracking algorithms and fluorescent staining of the secreted Psl exopolysaccharides (slime), they concluded that *P. aeruginosa* cells form cell clusters by depositing slime trails that influence the motility of their kin cells that encounter these trails, to follow and further strengthen the trails. These processes results in a positive feedback loop reinforcing the trails. Our study shows that *M. xanthus* cells use a similar mechanism to form aligned cell clusters. Furthermore, our results show that differences in surface motility mechanisms (e.g., reversals or the ability to follow trails) lead to distinct cell-clustering behaviors. These distinctions can be used to identify the nature of cell motility from snapshot images of bacteria for which direct observations on individual cells are difficult. Therefore, the mechanistic model of cell clustering and alignment developed here can be applicable to a wide class of bacteria displaying surface motility.

## Methods

## Agent-based simulation framework

### Biophysical model of *M. xanthus* cell

We have extended our previous biophysical model [20] for flexible *M. xanthus* cells to account for periodic cell reversals and slime-trail-following by cells. Brief description of the cell model along with changes introduced over the previous model is presented here. Refer to Balagam et al. [20] for additional details of our cell model. In the following sections bold letters indicate vectors and letters with a hat indicate unit vectors.

Each agent in this model consists of multiple segments enabling a realistic mechanical model of a single *M. xanthus* cell. We represent each agent as a connected string of $N(=7)$ nodes (Fig S7A). The first ($i=1$) and the last ($i=7$) nodes of the agent are designated as head and tail nodes respectively. Neighbor nodes are joined by rotational joints consisting of linear and angular springs. Linear springs (spring constant, $k_l$) between nodes resist elongation or compression to keep the agent length constant. Angular springs (spring constant, $k_b$) resist bending of the agent from straight line position and thus simulate elastic nature of cell bending.

For simplicity, we only implement gliding (A) motility of *M. xanthus* cells in our model. For this motility we use the distributed force generation along cell length through multiple motor protein complexes as indicated by recent models [25-28]. Thus, agents move forward through propulsion forces ($\boldsymbol{F}_p$) acting on nodes (except at tail node) tangential to the agent towards next ($i-1$) node in the current cell travel direction. Direction of propulsive force on the head node ($\boldsymbol{F}_{p,1}$) is influenced by other contributing factors (e.g. slime-trail following, random-turning noise etc. – explained below), in absence of which acts in a vector direction from its previous node ($i=2$) to the head node ($i=1$). We keep the magnitude of propulsive force ($F_{p,i} = F_T/(N-1)$, where $F_T$ is total propulsive force per cell) on each node equal. Viscous drag forces ($\boldsymbol{F}_d$) arising from the surrounding fluid act on nodes opposing their motion and are proportional to the node velocities (with proportionality coefficient/drag coefficient, $c$). Adhesive attachments to underlying substrate at nodes (except at the tail node) resist lateral displacement of the nodes during collisions. These attachments are modeled as linear springs (spring constant $k_a$) and are detached at a maximum breaking force $F_{a,max}$ ($=50\ pN$ [20]). These attachments represent the adhesion complexes in focal adhesion model of gliding motility in *M. xanthus* [26].

At low densities, *M. xanthus* cells are known to move in a monolayer. Therefore, collisions among agents are resolved by applying appropriate forces on nodes that keep agents from overlapping. Additionally, we employ appropriate forces on agent nodes to simulate periodic reversals of cells, noise in cell travel direction, and slime-trail-following by cells. Implementation details of these processes in our model are presented below.

**Periodic cell reversals**

*M. xanthus* cells periodically reverse their travel direction (mean reversal period = 8 min [4]) by switching the roles of its head and tail parts [41,42]. We mimic this behavior in our model by renumbering nodes in reverse order i.e., switching the roles of head and tail nodes at each reversal event and as a result the direction of propulsive force on agent nodes are rotated 180°. Reversals in agents are triggered asynchronously by an internal timer expiring at the end of the reversal period ($\tau_r$). This timer is reset to zero at each reversal event. During initialization, each agent's reversal timer is initialized randomly between $[0, \tau_r]$. For all simulations shown here reversals are perfectly periodic i.e., no noise in $\tau_r$ is introduced. However, introducing noise in reversals does not affect our conclusions (data not shown).

**Noise in cell travel direction**

*M. xanthus* cells exhibit random turns during movement on solid surfaces [20]. What triggers this random change in cell travel direction is not known. We introduce these random cell turns in our model by altering the direction of propulsive force on agents' head node. For simplicity, we only introduce a constant amount of noise in our model. Agents in our model change their travel direction during turn events that are activated asynchronously. During a turn event, we rotate the direction of the propulsion force on an agent's head node by 90° either clockwise or anti-clockwise direction chosen randomly (Fig S7C). Each turn event lasts for a fixed time interval (1 min). Similar to periodic reversals, turns events in each agent are activated through an internal timer, expiring after a fixed amount of time ($\tau_t = 5 \min$). During initialization, each agent's turn event timer is initialized randomly between $[0, \tau_t]$

**Slime-trail-following by cells**

The exact mechanism for slime-trail-following by *M. xanthus* cells is currently not known. It is possible that slime tracking by a cell is facilitated by attaching the type IV pili at the leading pole of the cell to the slime deposited on the substrate [43]. Retraction of the pili inward causes the cell to reorient towards the nearest slime trail. Alternatively, slime trails may provide low resistance (drag) paths compared to the slime-free areas and thus allow the cells slip into these paths when they cross these slime trails.

We employ a phenomenological approach for slime-trail-following in our model where we gradually change the direction of propulsive force ($\boldsymbol{F}_{p,1}$, Fig S7D) on an agent's leading node parallel to the slime-trail it is crossing. Here, we assume that cells actively seek slime rich regions on the substrate. Thus, we model a slime field covering the entire simulation region that tracks the amount of slime at each position. This slime field is divided into a square grid area with grid width equal to the cell width ($W_c$). Each agent secretes slime at a constant rate ($S_r$) as it moves forward, that is deposited into the underlying slime field grid elements. Slime exponentially degrades (or dries) in each grid element ($dS/dt = -k_d S$, where $k_d$ is the degradation constant). We assume that cells can only track wet slime (threshold slime detection limit = 1% of original deposit volume). Consecutive grid elements with wet slime represent a slime trail in our model.

Propulsive force on the head node ($F_{p,1}$) of an agent is influenced by the presence of nearby slime trails (Fig S7D, left). When an agent encounters a slime-trail, total propulsive force on its head node is rotated with its magnitude preserved and the rotation amount is a function of slime concentration. To implement this we split $F_{p,1}$ into two components: one in current head node direction ($F_c$) and another parallel to the slime-trail ($F_s$). The magnitude of force in slime direction is proportional to the fraction of slime remaining in the grid element whereas $F_c$ is computed to keep the magnitude of propulsion force, $F_T/(N-1)$, constant.

$$F_s = \left[\varepsilon_s \left(\frac{F_T}{N-1}\right)\left(\frac{S}{S_0}\right)\right]\hat{e}_s$$

$$F_c = \left[\frac{F_T}{N-1} - |F_s|\right]\hat{e}_h$$

Here $\varepsilon_s$ is slime effectiveness factor, $F_T$ – total propulsive force per cell, $N$ – number of nodes per cell, $S$ – volume of wet slime in grid element, $S_0$ – initial volume of wet slime, $\hat{e}_s, \hat{e}_h$ – unit vectors in direction of slime-trail and head node respectively.

We determine the direction of the dominant slime-trail ($\hat{e}_s$) using the following procedure (adapted from Hendrata et al.[44]). A semi-circular region, radius equal to half the cell length, in front of each cell's head node is designated as slime search region (Fig S7D, right). This semi-circle area is divided into 5 sectors (bins) and the total slime volume in each bin and the maximum slime volume ($S_{max}$) among the 5 bins are calculated. Finally, we estimate the slime-trail direction as the vector along the center line of the bin (sector) with at least 80% $S_{max}$ slime volume and has least angle deviation ($\Delta\theta_s$) from current head node direction ($\hat{e}_h$). If two bins (on opposite sides of $\hat{e}_h$) satisfy the above condition, then we chose either bin randomly.

Slime-trail length ($L_s$) is estimated as the distance travelled by an agent within the time slime deposited at a grid element degrades below a threshold volume ($S_{thr} = 0.01$). We assume that slime degrades exponentially with time (rate constant $k_d$). So the amount of slime deposited in a grid element of width ($W_c$) by the time ($\tau_1 = w_c / v_c$; $v_c$ - mean cell speed) an agent crosses the grid element is $S_{\tau_1} = (1 - e^{-k_d \tau_1}) S_r / k_d$. And the time ($t_{thr}$) required to degrade this initial deposited slime volume ($S_{\tau_1}$) below the threshold volume is $t_{thr} = \ln(S_{\tau_1} / S_{thr}) / k_d$. Finally, slime-trail length of is calculated as $L_s = t_{thr} v_c$.

To test the robustness of our results using slime-trail-following mechanism we have varied the length of the slime-trail ($L_s$) produced by an agent. For this, we have multiplied production rate of slime ($S_r$) from an agent and slime degradation constant ($k_d$) with same factor so that the net volume of slime in the simulation region remains constant.

**Lateral cell adhesions**

To simulate adhesive interactions between agents (used only for simulations in Fig S5), we apply lateral adhesive forces ($F_{adh}$) on nodes of neighboring agents that are closer than specific threshold distance $d_{thr} (= 0.75 \, \mu m)$ (Fig S7E). Adhesive force on each node is calculated using the following equation.

$$F_{adh} = \begin{cases} 0 & d_\perp \geq d_{thr} \\ \left( \dfrac{d_\perp - W_c}{d_{thr} - W_c} \right) k_{adh} \dfrac{F_T}{N} & W_c \leq d_\perp < d_{thr} \end{cases}$$

Here $d_\perp$ is the perpendicular distance between the nodes of neighboring agents. These adhesive forces are applied on each node normal to the direction of propulsive force ($\hat{e}_{n,i}$) towards its neighbor agent nodes.

**Simulation procedure**

We study the clustering behavior of cells by simulating mechanical interactions among large number ($M$) of agents on a 2D simulation region with periodic boundary conditions in an agent-based-simulation (ABS) framework. Flow chart for our simulation procedure is shown in Fig S8.

We initialize agents one by one on a square simulation region (dimension $L_{sim}$) over few initial time steps until desired cell density ($\eta$) is reached. Agents are initialized in random positions over the simulation region with their orientations ($\theta$) chosen randomly in the range $[0, 2\pi]$. Here, an agent orientation is defined as the angle made by the vector pointing from its tail node to head node with X-axis (Fig S7B). Agent nodes are initialized in straight-line configuration.

During initialization, agent configurations that overlap with existing agents are rejected. After initialization the head node for each agent is chosen between its two end-nodes ($i = 1,7$) with 50% probability.

At each time step of simulation, agents move according to the various forces (see Fig S8) acting on their nodes. Changes in node positions and velocities are obtained by integrating the equations of motion based on Newton's laws. We use Box2D physics library [29] for solving the equations of motion and for effective collision resolution. Snapshots of the simulation region, orientation of each agent along with its node positions are recorded at 1 min time interval for later analysis.

Simulations are implemented in Java programming language with a Java port of Box2D library (http://www.jbox2d.org/). Parameters of the simulation are shown in Supp. Text. Other parameters of the model are same as in Balagam et al.[20]. Each simulation is run for 180 min.

## Acknowledgements

We thank Joshua Shaevitz and his lab (Princeton University) for sharing their data from Ref. [17].

## Funding

The research is supported by NSF awards MCB-1411780, PHY- 1427654 and K2I 2014/15 OG HPC workshop graduate fellowship.

# References

1. Kaiser D (2003) Coupling cell movement to multicellular development in myxobacteria. Nature reviews Microbiology 1: 45-54.
2. Whitworth DE, American Society for Microbiology. (2008) Myxobacteria : multicellularity and differentiation. Washington, DC: ASM Press. xvii, 520 p., 512 p. of plates p.
3. Kaiser D (2008) Myxococcus-from single-cell polarity to complex multicellular patterns. Annual review of genetics 42: 109-130.
4. Berleman JE, Chumley T, Cheung P, Kirby JR (2006) Rippling is a predatory behavior in Myxococcus xanthus. Journal of bacteriology 188: 5888-5895.
5. Igoshin OA, Mogilner A, Welch RD, Kaiser D, Oster G (2001) Pattern formation and traveling waves in myxobacteria: theory and modeling. Proc Natl Acad Sci U S A 98: 14913-14918.
6. Zhang H, Vaksman Z, Litwin DB, Shi P, Kaplan HB, et al. (2012) The mechanistic basis of Myxococcus xanthus rippling behavior and its physiological role during predation. PLoS computational biology 8: e1002715.
7. Kaiser D, Welch R (2004) Dynamics of fruiting body morphogenesis. Journal of bacteriology 186: 919-927.
8. Kuner JM, Kaiser D (1982) Fruiting body morphogenesis in submerged cultures of Myxococcus xanthus. Journal of bacteriology 151: 458-461.
9. Sliusarenko O, Zusman DR, Oster G (2007) Aggregation during fruiting body formation in Myxococcus xanthus is driven by reducing cell movement. Journal of bacteriology 189: 611-619.
10. Igoshin OA, Kaiser D, Oster G (2004) Breaking symmetry in myxobacteria. Current biology : CB 14: R459-462.
11. Wall D, Kaiser D (1998) Alignment enhances the cell-to-cell transfer of pilus phenotype. P Natl Acad Sci USA 95: 3054-3058.
12. Welch R, Kaiser D (2001) Cell behavior in traveling wave patterns of myxobacteria. Proc Natl Acad Sci U S A 98: 14907-14912.
13. Pelling AE, Li Y, Cross SE, Castaneda S, Shi W, et al. (2006) Self-organized and highly ordered domain structures within swarms of Myxococcus xanthus. Cell Motil Cytoskel 63: 141-148.
14. O'Connor KA, Zusman DR (1989) Patterns of cellular interactions during fruiting-body formation in Myxococcus xanthus. Journal of bacteriology 171: 6013-6024.
15. Kim SK, Kaiser D (1990) Cell Alignment Required in Differentiation of Myxococcus-Xanthus. Science 249: 926-928.
16. Starruss J, Peruani F, Jakovljevic V, Sogaard-Andersen L, Deutsch A, et al. (2012) Pattern-formation mechanisms in motility mutants of Myxococcus xanthus. Interface Focus 2: 774-785.
17. Thutupalli S, Sun M, Bunyak F, Palaniappan K, Shaevitz JW (2014) Phase transitions during fruiting body formation in Myxococcus xanthus. arxivorg. arxiv.org.
18. Harvey CW, Alber M, Tsimring LS, Aranson IS (2013) Continuum modeling of clustering of myxobacteria. New journal of physics 15.
19. Peruani F, Starruss J, Jakovljevic V, Sogaard-Andersen L, Deutsch A, et al. (2012) Collective Motion and Nonequilibrium Cluster Formation in Colonies of Gliding Bacteria. Phys Rev Lett 108.

20. Balagam R, Litwin DB, Czerwinski F, Sun M, Kaplan HB, et al. (2014) Myxococcus xanthus gliding motors are elastically coupled to the substrate as predicted by the focal adhesion model of gliding motility. PLoS computational biology 10: e1003619.
21. Peruani F, Deutsch A, Bar M (2006) Nonequilibrium clustering of self-propelled rods. Physical review E, Statistical, nonlinear, and soft matter physics 74: 030904.
22. McCandlish SR, Baskaran A, Hagan MF (2012) Spontaneous segregation of self-propelled particles with different motilities. Soft Matter 8: 2527-2534.
23. Peruani F, Schimansky-Geier L, Bar M (2010) Cluster dynamics and cluster size distributions in systems of self-propelled particles. Eur Phys J-Spec Top 191: 173-185.
24. Yang Y, Marceau V, Gompper G (2010) Swarm behavior of self-propelled rods and swimming flagella. Physical review E, Statistical, nonlinear, and soft matter physics 82: 031904.
25. Nan B, Zusman DR (2011) Uncovering the mystery of gliding motility in the myxobacteria. Annual review of genetics 45: 21-39.
26. Sun M, Wartel M, Cascales E, Shaevitz JW, Mignot T (2011) Motor-driven intracellular transport powers bacterial gliding motility. Proc Natl Acad Sci U S A 108: 7559-7564.
27. Mignot T, Shaevitz JW, Hartzell PL, Zusman DR (2007) Evidence that focal adhesion complexes power bacterial gliding motility. Science 315: 853-856.
28. Sliusarenko O, Zusman DR, Oster G (2007) The motors powering A-motility in Myxococcus xanthus are distributed along the cell body. Journal of bacteriology 189: 7920-7921.
29. Catto E (2012) Box2D - A 2D physics engine for games http://box2d.org/.
30. Ginelli F, Peruani F, Bar M, Chate H (2010) Large-scale collective properties of self-propelled rods. Phys Rev Lett 104: 184502.
31. Janulevicius A, van Loosdrecht MC, Simone A, Picioreanu C (2010) Cell flexibility affects the alignment of model myxobacteria. Biophysical journal 99: 3129-3138.
32. Berleman JE, Scott J, Chumley T, Kirby JR (2008) Predataxis behavior in Myxococcus xanthus. Proc Natl Acad Sci U S A 105: 17127-17132.
33. Shimkets LJ (1986) Role of cell cohesion in Myxococcus xanthus fruiting body formation. Journal of bacteriology 166: 842-848.
34. Buttinoni I, Bialke J, Kummel F, Lowen H, Bechinger C, et al. (2013) Dynamical clustering and phase separation in suspensions of self-propelled colloidal particles. Phys Rev Lett 110: 238301.
35. Shimkets LJ (1986) Correlation of energy-dependent cell cohesion with social motility in Myxococcus xanthus. Journal of bacteriology 166: 837-841.
36. Behmlander RM, Dworkin M (1994) Biochemical and structural analyses of the extracellular matrix fibrils of Myxococcus xanthus. Journal of bacteriology 176: 6295-6303.
37. Wolgemuth C, Hoiczyk E, Kaiser D, Oster G (2002) How myxobacteria glide. Current biology : CB 12: 369-377.
38. Burchard RP (1982) Trail following by gliding bacteria. Journal of bacteriology 152: 495-501.
39. Janulevicius A, van Loosdrecht M, Picioreanu C (2015) Short-range guiding can result in the formation of circular aggregates in myxobacteria populations. PLoS computational biology 11: e1004213.
40. Zhao K, Tseng BS, Beckerman B, Jin F, Gibiansky ML, et al. (2013) Psl trails guide exploration and microcolony formation in Pseudomonas aeruginosa biofilms. Nature 497: 388-391.

41. Wu Y, Kaiser AD, Jiang Y, Alber MS (2009) Periodic reversal of direction allows Myxobacteria to swarm. Proc Natl Acad Sci U S A 106: 1222-1227.
42. Yu R, Kaiser D (2007) Gliding motility and polarized slime secretion. Molecular microbiology 63: 454-467.
43. Li Y, Sun H, Ma X, Lu A, Lux R, et al. (2003) Extracellular polysaccharides mediate pilus retraction during social motility of Myxococcus xanthus. Proc Natl Acad Sci U S A 100: 5443-5448.
44. Hendrata M, Yang Z, Lux R, Shi W (2011) Experimentally guided computational model discovers important elements for social behavior in myxobacteria. Plos One 6: e22169.

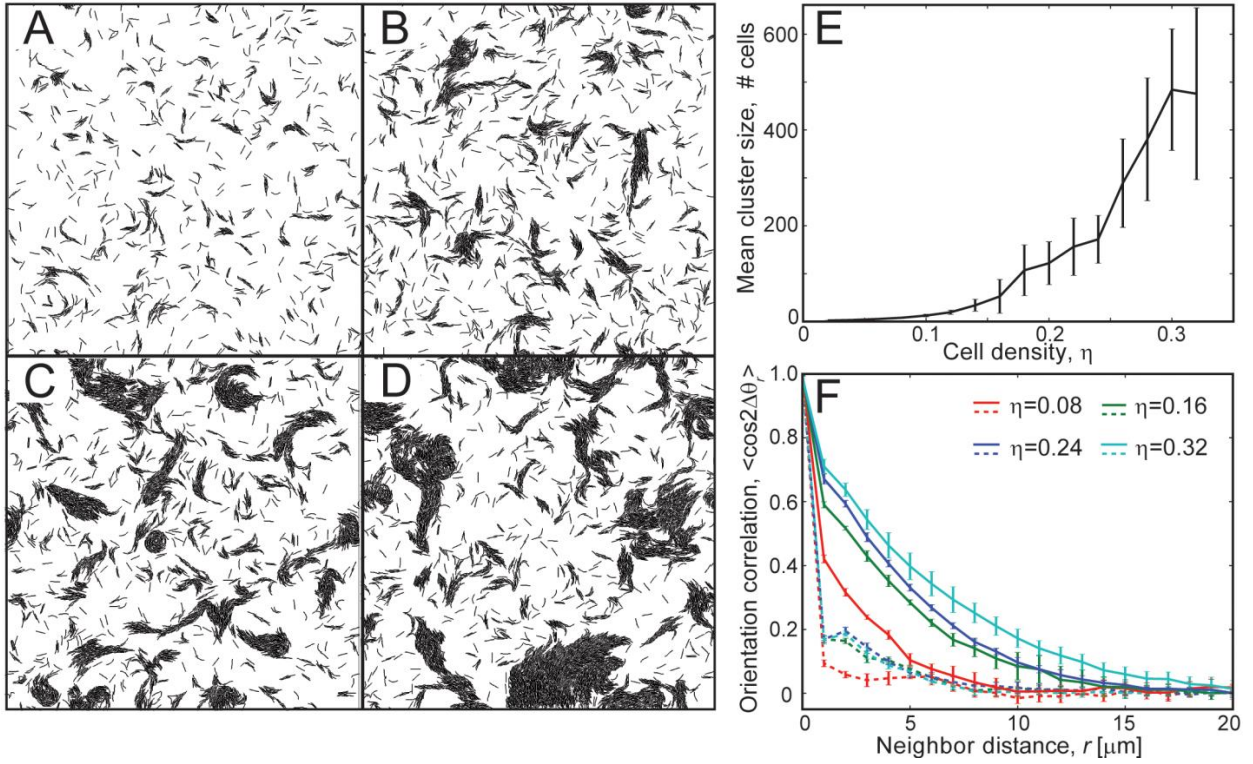

**Figure 1 Clustering behavior of non-reversing flexible agents in simulations**

(A-D) Snapshots of the simulation region at 180 min of simulation time for different cell densities, $\eta$. (A) $\eta = 0.08$, (B) $\eta = 0.16$, (C) $\eta = 0.24$, (D) $\eta = 0.32$. Flexible agents formed aligned clusters at moderate to high cell densities ($\eta \geq 0.16$). (E) Mean cluster sizes, $\langle m \rangle$, from simulation as a function of cell density, $\eta$. The error bars indicate the standard deviation in the data. The results are averaged over 5 independent simulation runs. The mean cluster sizes increased with increases in cell density. (F) Orientation correlation $\langle \cos 2\Delta\theta_r \rangle$ among cells as a function of neighbor cell distance, $r$. $\Delta\theta_r$ is the angle deviation between orientations ($\theta$) of a pair of neighbor cells separated by a distance $r$. Orientation correlation ($\cos 2\Delta\theta_r$) values from all cell pairs are binned based on $r$ (bin width = 1 $\mu m$) and averaged. Dashed and solid lines represent orientation correlation values at 1 min and 180 min of simulation time, respectively. Agents in clusters showed higher neighbor alignment at larger distances compared to the initial randomly oriented cells. Furthermore, the alignment increases with increases in cell density.

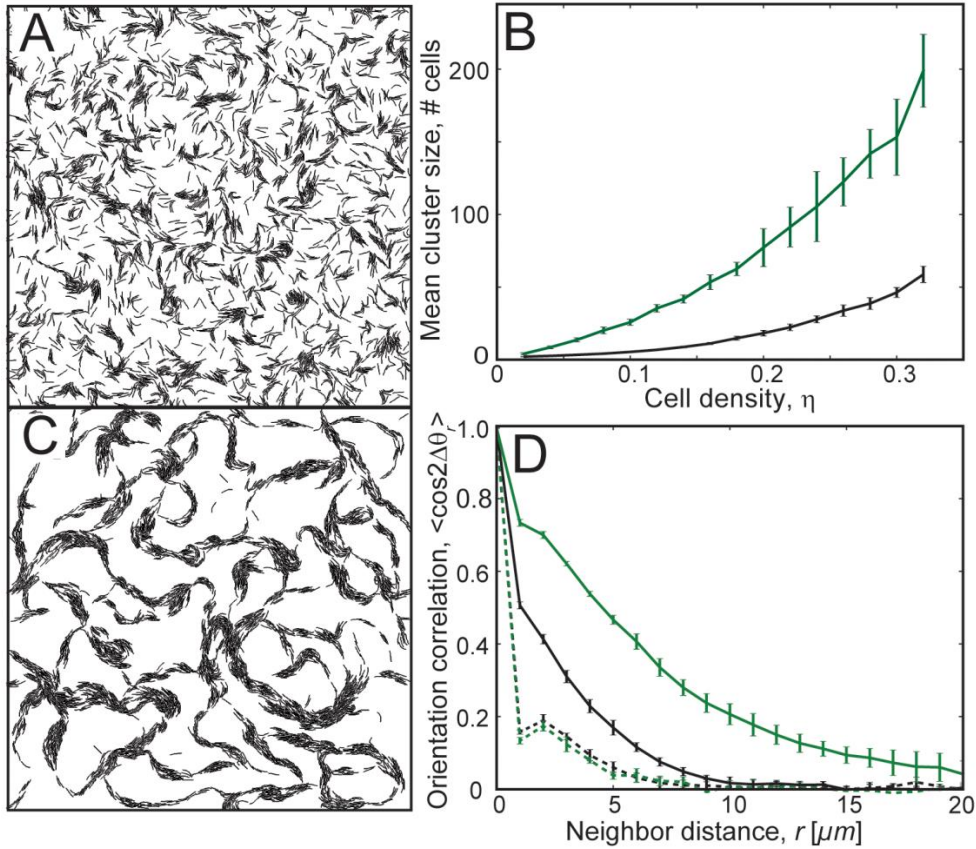

**Figure 2 Clustering behavior of periodically reversing agents in simulations**

(A) Snapshot of the simulation with periodically reversing agents ($\eta = 0.24$) at 180 min of simulation time. Reversing agents did not show significant clustering. (B) Mean cluster sizes, $\langle m \rangle$, in simulation as a function of cell density, $\eta$, for agents following slime trails (green line) and agents without slime trails (black line). Agents following slime trails showed a significant increase in mean cluster size compared to agents without slime-trail-following. (C) Snapshot of the simulation for periodically reversing cells with the slime-trail-following mechanism ($\eta = 0.24, L_s = 11\,\mu m, \varepsilon_s = 1.0$, refer to Methods for details) at 180 min of simulation time. Agents show improved clustering compared to those without the slime-trail-following mechanism. (D) Orientation correlation $\langle \cos 2\Delta\theta_r \rangle$ among agents for reversing cells (black) and reversing cells with the slime-trail-following mechanism (green). Dashed and solid lines are orientation correlation values at 1 min and 180 min of simulation time, respectively. Orientation correlation with neighbors improved for larger neighbor distances with the slime-trail-following mechanism.

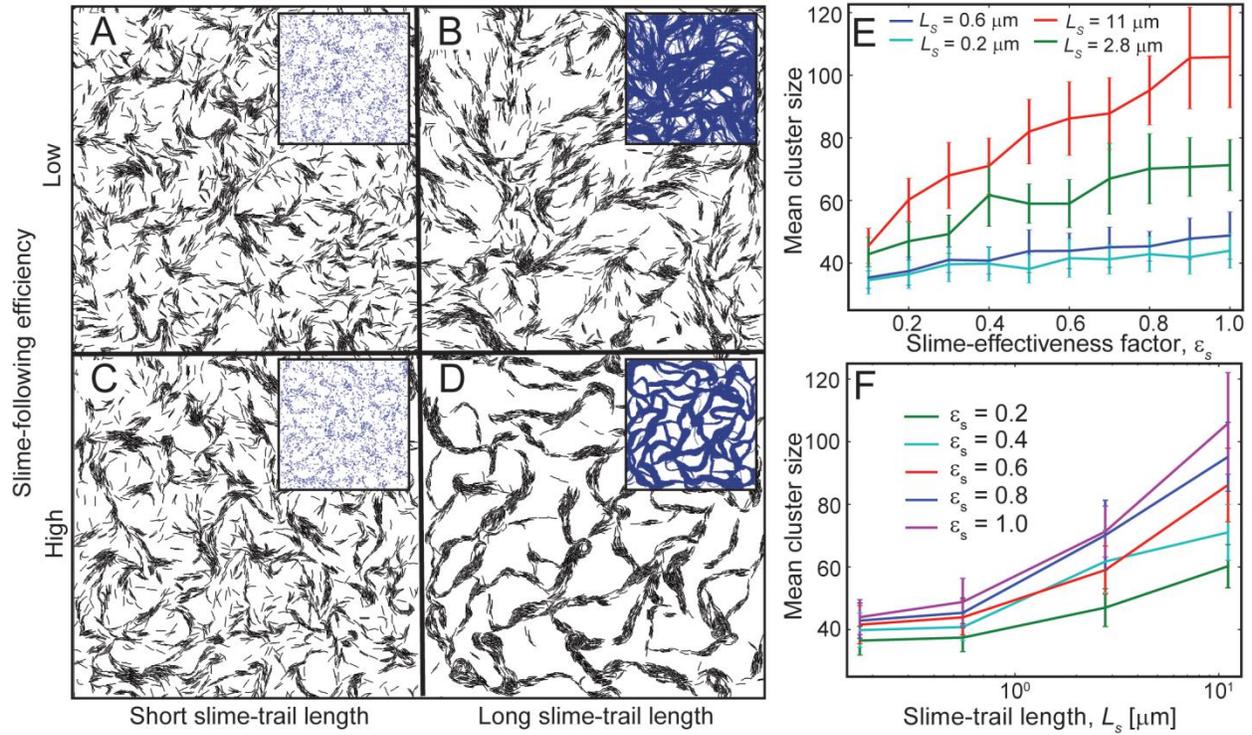

**Figure 3 Robustness of the slime-trail-following mechanism for cell clustering**

(A-D) Snapshots of simulations showing agent clustering behavior ($\eta = 0.24$) for variation in the slime effectiveness value and slime trail length at 180 min of simulation time. Only agents with high slime-trail-following efficiency and long slime trails show significant clustering behavior (D). Inset figures show the slime distribution in the simulation region. The mean cluster sizes in the simulations (E) as a function of the slime effectiveness factor, $\varepsilon_s$ for different slime trail lengths and (F) as a function of the slime trail length, $L_s$, for different slime effectiveness factor values. Cell clustering improved with increases in the slime effectiveness factor (E), provided the slime trails are sufficiently long, and with increases in the slime trail length (F).

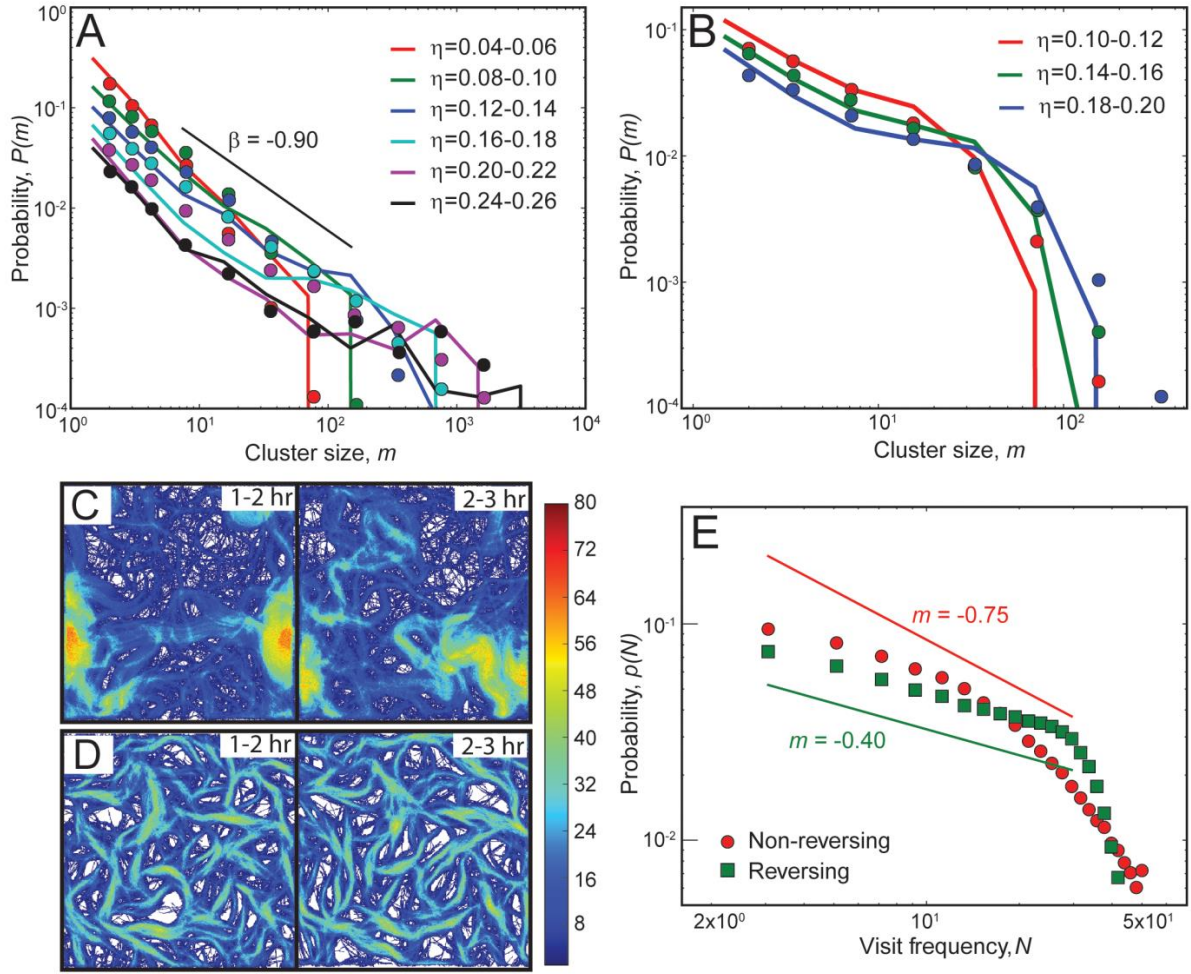

**Figure 4 Comparison of cell clustering behavior in simulations with experiments**

(A-B) Comparison of cluster size distributions (CSD) from simulations (lines) with experimental data (symbols, digitized from Starruß et al. [16]) for non-reversing (A) and reversing (B) cells. Probability, $p(m)$, of finding a cell in a cluster is plotted as a function of the cluster size $m$. We use different sets of slime-trail-following mechanism parameters for non-reversing ($L_s = 0.6\,\mu m, \varepsilon_s = 0.5$) and reversing ($L_s = 11\,\mu m, \varepsilon_s = 0.2$) agents. CSD results from simulations show a similar trend to that of the experimental data. (A) Non-reversing cells show a power-law-like CSD, whereas reversing cells show a monotonically decreasing CSD (B). (C-D) Heat maps of cell visit frequencies over the simulation region for 2 consecutive hours ($\eta = 0.24$). The color bar represents the number of cell visits per hour at a particular location. Non-reversing cells show a dynamic cluster pattern with changes in cell traces (C), whereas reversing cells show a static cluster pattern with the pattern of cell traces remaining approximately the same over time (D). (E) Probability of cell visits, $p(N)$, as a function of visit frequency, $N$, for non-reversing (red) and reversing cells (green) over a 1-hr simulation time (120-180 min). Reversing cells show a large fraction of sites with high visit frequencies compared to non-reversing cells.

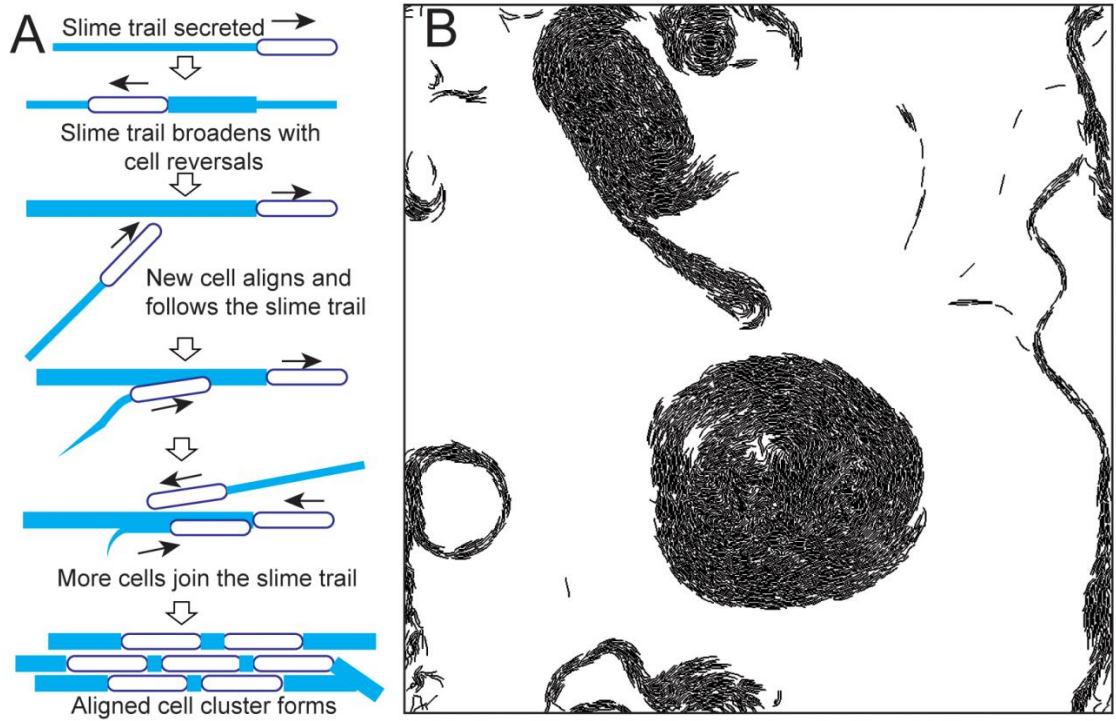

**Figure 5** (A) Hypothetical mechanism of cell clustering through slime-trail-following in reversing *M. xanthus* cells. (B) Circular cell aggregates observed in simulation for non-reversing agents with the slime-trail-following mechanism ($\eta = 0.24, L_s = 11\,\mu m, \varepsilon_s = 1.0$)

# Supplementary Text

## Quantitative measurements of cell clustering

### Cluster size distribution (CSD)

We identify the clusters in the simulation region using a density based clustering algorithm (DBSCAN [SR, SR6]) applied on agent node positions. This algorithm identifies groups of nodes that exceed given density threshold and classifies them as separate clusters. We chose the parameters of the algorithm (minimum number of nodes to form a cluster, $k = 21$ i.e., 3 cells, and the minimum neighbor distance, $d_{min} = 2$) that resulted in good separation (visually) between individual clusters. Using small neighbor distance ($d_{min} \leq 2$) values in this algorithm resulted in large clusters that are actually multiple separate clusters connected by a narrow streams of agents. So we used minimum neighbor distance as $d_{min} = 2\,\mu m$ and later processed the individual clusters to include short distance ($\leq 2\,\mu m$) neighbor agents.

Next, we determine the agents belonging to each separate cluster of nodes identified by the algorithm. We process partial agents i.e., agents for which only fraction of their nodes are included into a cluster, to include all their nodes into the cluster. We further process the clusters to include all the nearest neighbor agents ($d_{min} < 0.75\,\mu m$) that are missed by the algorithm. We quantify the size of the clusters ($m$) by measuring the number of agents in each cluster. Snapshots of identified clusters (after processing) from simulation are shown in M1Fig.

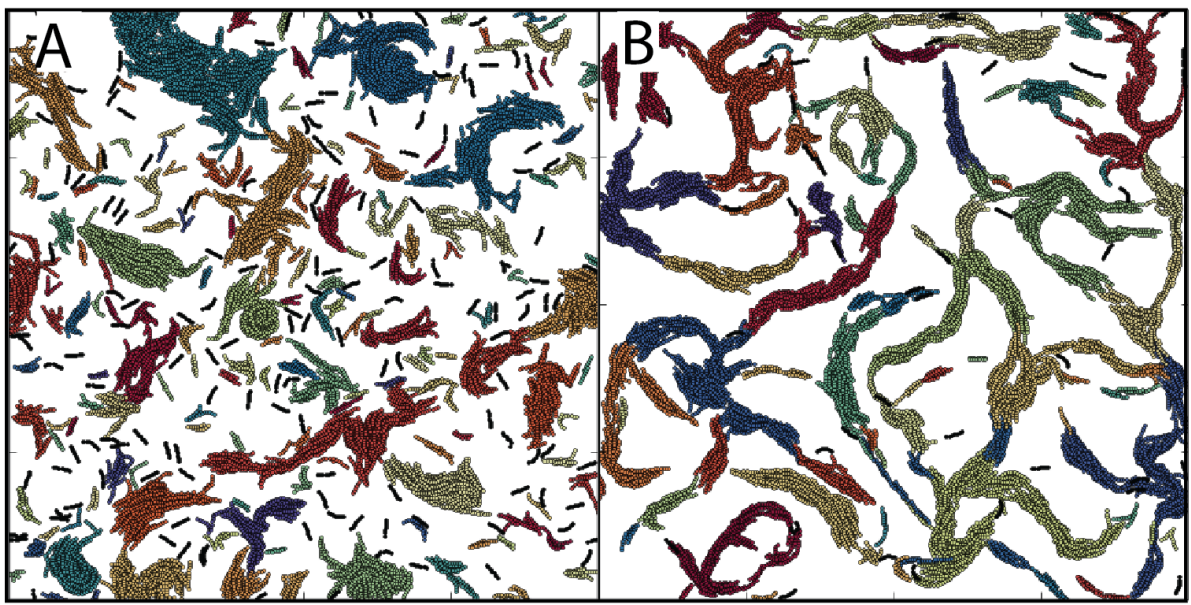

**Figure M1 Identifying cell clusters using DBSCAN algorithm** Different colors represent different clusters of agent nodes identified from their positions that exceed given density threshold. Snapshots of identified clusters for (A) Non-reversing cells without slime-trails (B) Reversing cells with slime-trail-following mechanism ($L_s = 11\,\mu m, \varepsilon_s = 1.0$) at 180 mins after the simulation started. Cell density $\eta = 0.24$

We quantify the cluster size distribution (CSD) by measuring the probability $p(m)$ of finding a cell in a cluster of size $m$. For this, we follow the procedure illustrated in Starruß et al.[16]. After identifying and processing the clusters from simulation, we obtain an array of various cluster sizes ($m_i$) at each time frame. These values ($m_i$) are converted into a normalized histogram ($f_i$) whose bin edges are chosen exponentially i.e., bin $i$ ($=1,2,...$) contains all agents that belong to clusters of size $10^{(i-1)dm} \leq m < 10^{i\,dm}$, where $dm = 0.33$. Thus $f_i$ values represent fraction of all agents found in cluster sizes represented by bin $i$. Finally, probabilities, $p_i$, are calculated by dividing the $f_i$ values with the corresponding bin width $db_i \left(=10^{i\,dm} - 10^{(i-1)dm}\right)$.

**Mean cluster size, ⟨m⟩**

Due to the sparse nature of cluster size distribution data from simulation, we calculate mean cluster size at each cell density ($\eta$) using data from multiple simulation runs and from multiple time points ($\tau_{ss,t}$) in each simulation run after CSD values reached steady state. We chose time points ($\tau_{ss,t}$) such that the CSD at these time points are sufficiently independent. For this we measure the auto-correlation between snapshots of cluster images from simulation as a function of time (M2 Fig). Image auto-correlation is calculated using Eqn. 1 where $C_I(t,t+\delta t)$ is the normalized cross correlation between snapshots ($f, g$) of the simulation at times $t$ and $t + \delta t$, $N_K$ is number of such image pairs. Normalized cross correlation between two images $f, g$ is calculated using Eqn. 2 where $n$ is the total number of pixels in the image, $f_{i,j}$ is grayscale intensity of pixel at position $i, j$ in image $f$, and $\bar{f}, \sigma_f$ are the average intensity and standard deviation in intensity of pixels in image $f$.

$$C(\delta t) = \frac{1}{N_K} \sum_{t=1}^{K} C_I(t, t+\delta t) \qquad (1)$$

$$C_I = \frac{1}{n} \sum_{i,j} \frac{\left(f_{i,j} - \bar{f}\right)\left(g_{i,j} - \bar{g}\right)}{\sigma_f \sigma_g} \qquad (2)$$

Auto-correlation values are measured for snapshots of simulation between 60 to 180 mins after initialization. From these results, we determined that correlation among cluster images dropped to low value (< 0.1) after 20 mins for both reversing and non-reversing cells (M2 Fig). Thus we take data from steady state time points ($\tau_{ss,t}$) separated by 20 mins as independent trials.

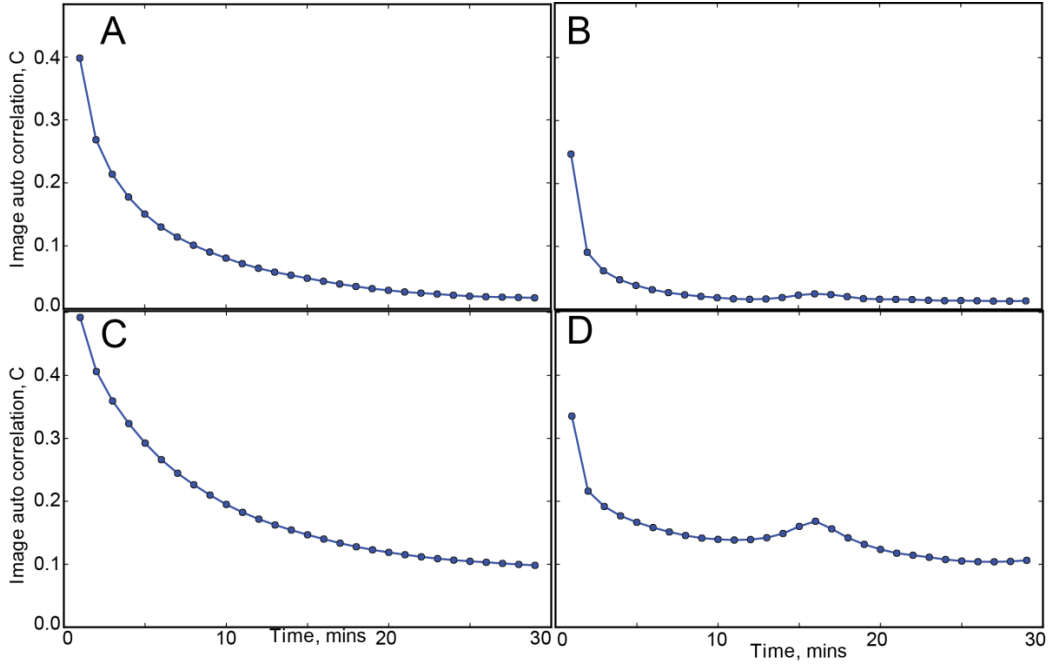

**Figure M2 Auto-correlation of cluster images with time**. Simulations with (A, C) non-reversing cells, (B, D) reversing cells. (C, D) Simulations with slime-trail-following mechanism with (C) model parameters $L_s = 0.6\,\mu m, \varepsilon_s = 0.5$ and (D) model parameters $L_s = 11\,\mu m, \varepsilon_s = 0.2$. All these simulations are performed at cell density $\eta = 0.24$. Correlation values are computed for snapshot images from 60 - 180 mins of the simulation time. Correlation among simulation images dropped to low value ($<0.1$) after 20 mins.

Mean cluster sizes at each time point ($\tau_{ss,t}$) is calculated using $\bar{m}_t = \sum_i p_i m_i$, where $p_i$ is the probability from CSD and $m_i$ is the average cluster size ($= 10^{(i-1)dm} + (10^{(i-1)dm} - 10^{i\,dm})/2$) of bin $i$. Finally the mean $\langle m \rangle$ and standard deviation in cluster sizes at each cell density is calculated by averaging data from all the steady state time points from multiple simulation runs ($n = 5$).

**Orientation correlation, $C(r)$**

Alignment among neighbor agents is quantified using orientation correlation function $C(r) = \langle \cos 2\Delta\theta_r \rangle$. Here $\Delta\theta_r$ is the angle deviation between orientations of a pair of neighbor cells whose center nodes are separated by a distance $r$ (Fig S7F). $\langle \ \rangle$ represents average over all cell pairs that are separated by a distance $r$.

**Table M1  Parameters used in simulation**

| Symbol | Description | Value |
|---|---|---|
| $L_{sim}$ | Dimension of square simulation region | $200 \, \mu m$ |
| $M$ ($\eta$) | Total number of agents (corresponding cell densities) | 246-3938 (0.02 – 0.32) |
| $\varepsilon_s$ | Slime effectiveness factor | 1.0 |
| $L_s$ | Slime-trail length | $11 \, \mu m$ |
| $k_d$ | Slime degradation constant | 1.0  1/min |
| $V_s$ | Slime production rate | 20 AU |
| $L_c$ | Agent length | $6.5 \, \mu m$  [6, SR3] |
| $W_c$ | Agent width | $0.5 \, \mu m$  [6, SR3] |
| $k_b$ | Angular spring constant/bending stiffness | $10^{-17} \, Nm$ [SR3-5] |
| $v_c$ | Mean speed of cell | $4 \, \mu m/min$  [41] |
| $t_{step}$ | Simulation time step | $5 \times 10^{-3}$ min |
| $k_a$ | Spring constant for cell-substrate adhesions | $100 \, pN/\mu m$ [20] |
| $F_T$ | Propulsive force per cell | 60 pN [20] |
| $\tau_r$ | Reversal period | 8 min [4] |

## Supplemental References

SR1. Ester M, Kriegel H-P, Sander J, Xu X. A density-based algorithm for discovering clusters in large spatial databases with noise 1996.
SR2. Pedregosa F, Varoquaux G, Gramfort A, Michel V, Thirion B, et al. (2011) Scikit-learn: Machine Learning in Python. J Mach Learn Res 12: 2825-2830.
SR3. Janulevicius A, van Loosdrecht MC, Simone A, Picioreanu C (2010) Cell flexibility affects the alignment of model myxobacteria. Biophysical journal 99: 3129-3138.
SR4. Wolgemuth CW (2005) Force and flexibility of flailing myxobacteria. Biophysical journal 89: 945-950.
SR5. Harvey CW, Morcos F, Sweet CR, Kaiser D, Chatterjee S, et al. (2011) Study of elastic collisions of Myxococcus xanthus in swarms. Physical biology 8: 026016.

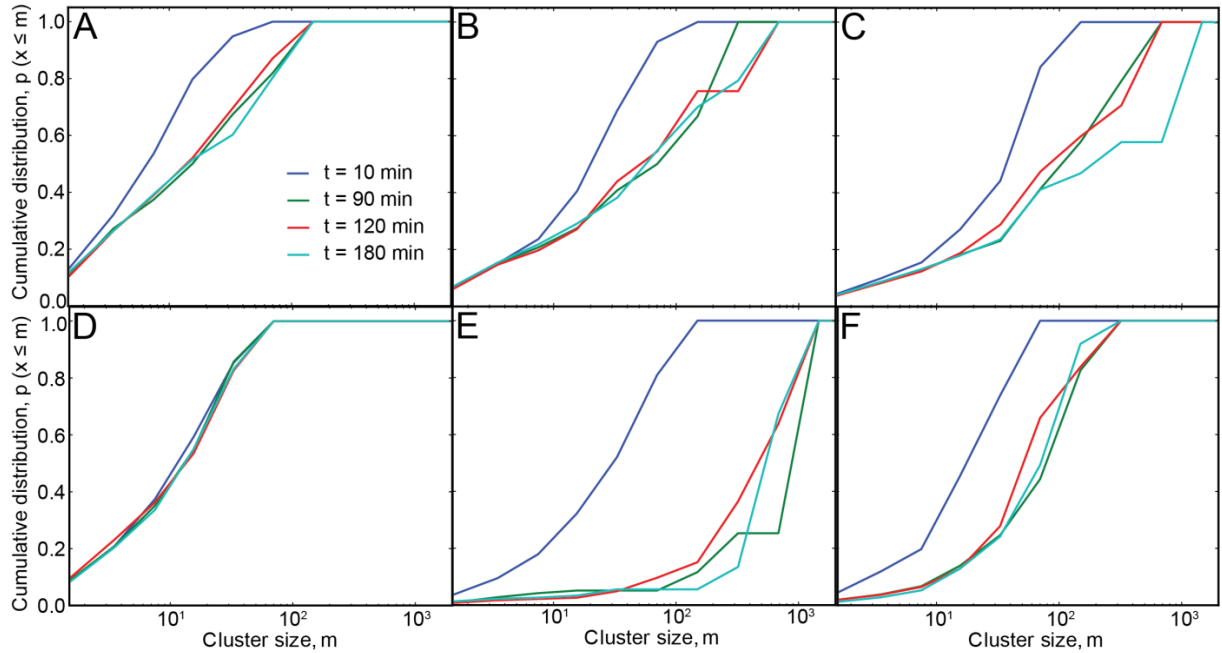

**Figure S1 Evolution of cumulative cluster size distribution (CSD) with time for different model parameters** Non-reversing agents with cell densities (A) $\eta = 0.08$ (B) $\eta = 0.16$ (C) $\eta = 0.24$ (D) $\eta = 0.32$ (E) Reversing agents with cell density $\eta = 0.24$ (F) Non-reversing agents following slime-trails, $\eta = 0.24$ (G) Reversing agents following slime-trails, $\eta = 0.24$.

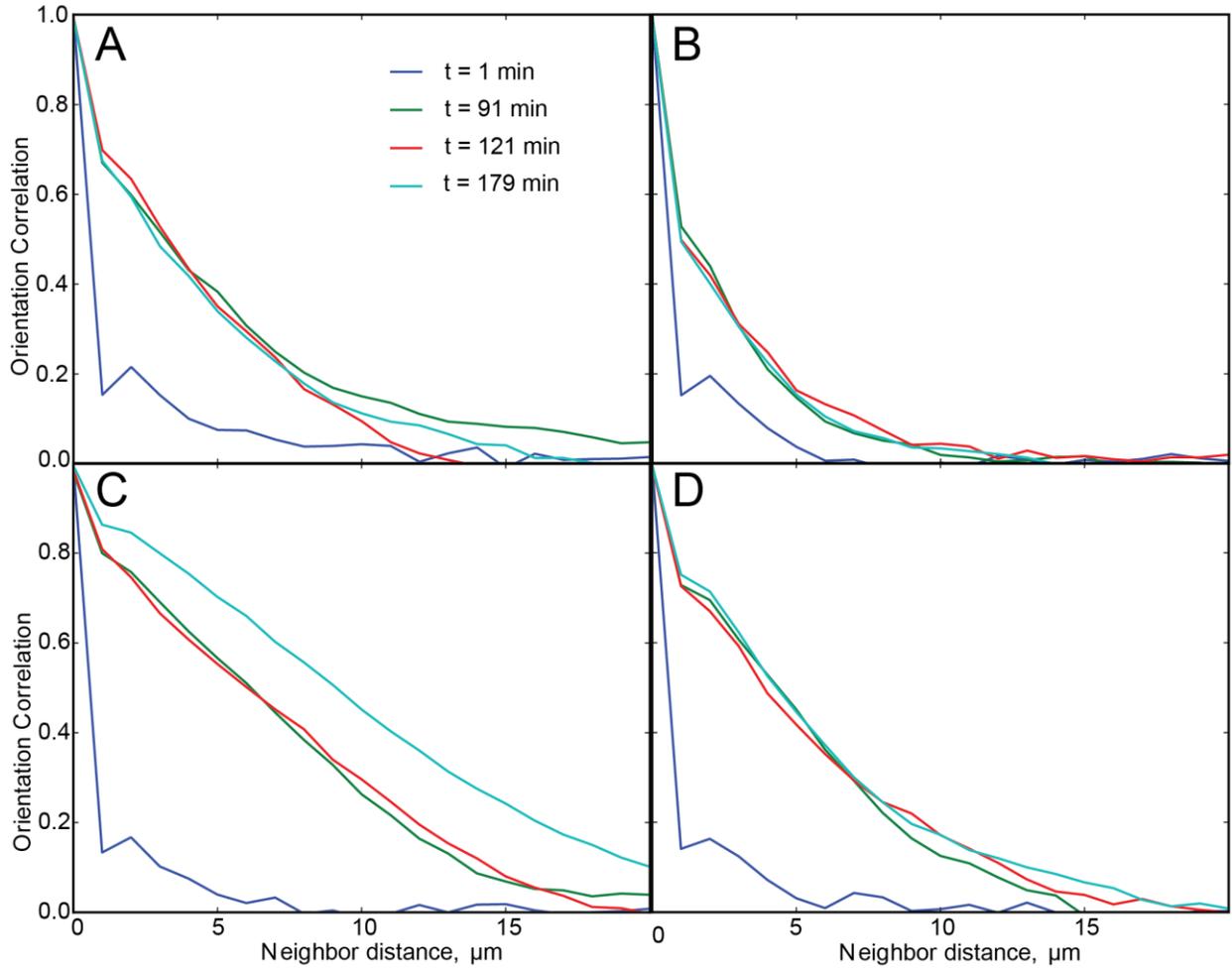

**Figure S2 Evolution of orientation correlation among cells with time** (A, C) Non-reversing cells (B, D) Reversing cells (C, D) cells following slime-trails. All simulations performed at cell density $\eta = 0.24$.

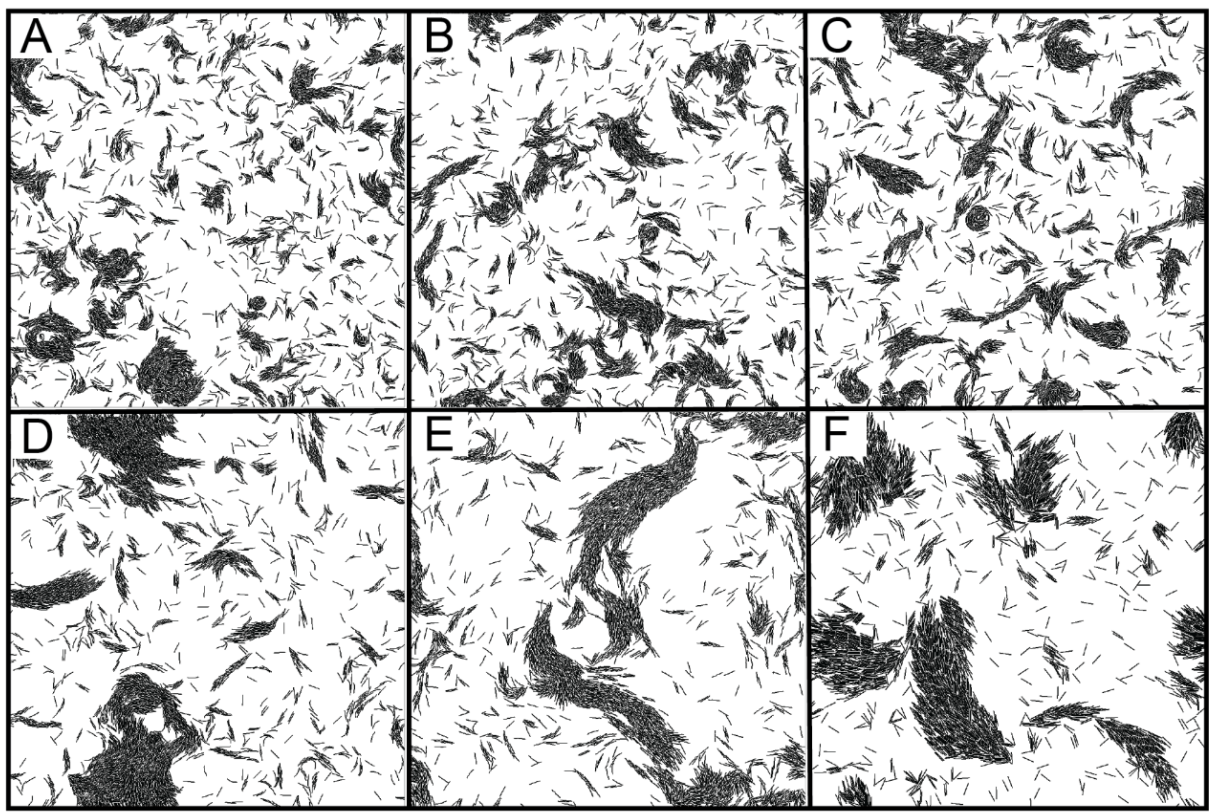

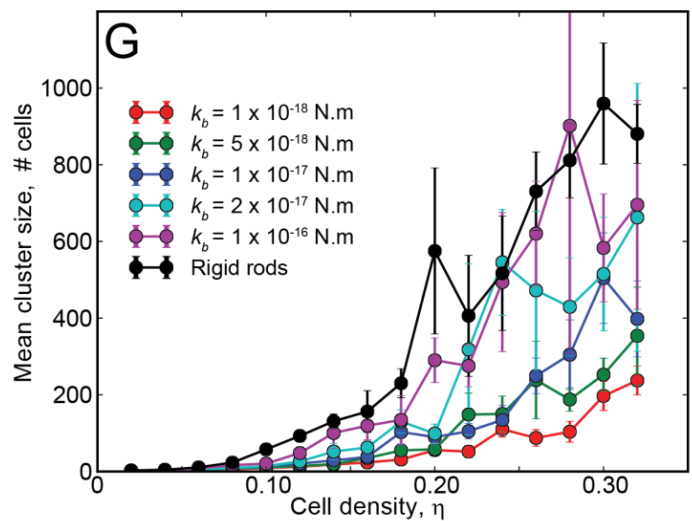

**Figure S3 Clustering behavior of non-reversing agents for variation in cell flexibility** (A-F) Snapshots of cell clusters after 180 min of simulation (cell density, $\eta = 0.24$) with bending stiffness ($k_b$) values (A) $10^{-18}\,N.m$ (B) $5\times10^{-18}\,N.m$ (C) $10^{-17}\,N.m$ (D) $2\times10^{-17}\,N.m$ (E) $10^{-16}\,N.m$ (F) Rigid rods (G) Mean cluster sizes in simulation as a function of cell density ($\eta$) for different cell bending stiffness values.

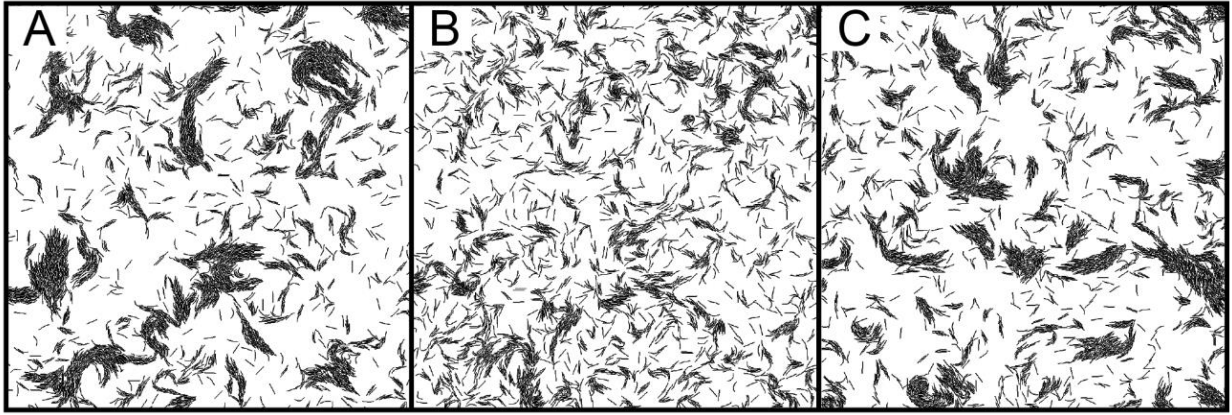

**Figure S4 Clustering behavior of cells with with turning on/off cell reversals** Snapshots of simulation at (A) 90 min (B) 120 min (C) 150 min. Cell reversals were turned-off for the first 90 min of simulation and thereafter are turned-on from 90 to 120 min with reversal period = 8 min. Reversals are turned-off again at 120 min. Cell clusters formed by simulating non-reversing cells for first 90 min (A) are quickly, within 30 min destroyed by cell reversals (B). Suppression of reversals restored clustering of cells after another 30 min (C).

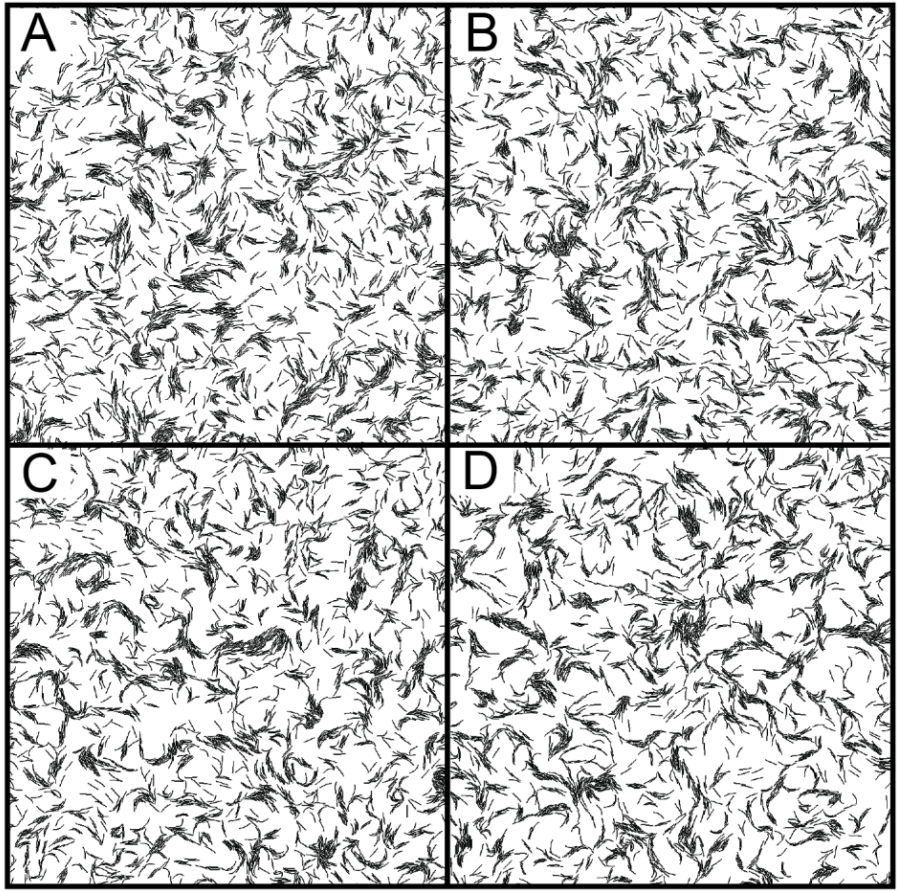

**Figure S5 Clustering behavior of reversing-cells with lateral cell adhesions** Snapshots of simulation at 180 min for different lateral adhesion force values. Adhesion force per cell (A) $F_{adh} = 0 \, pN$ (B) $F_{adh} = 30 \, pN$ (C) $F_{adh} = 60 \, pN$ (D) $F_{adh} = 120 \, pN$.

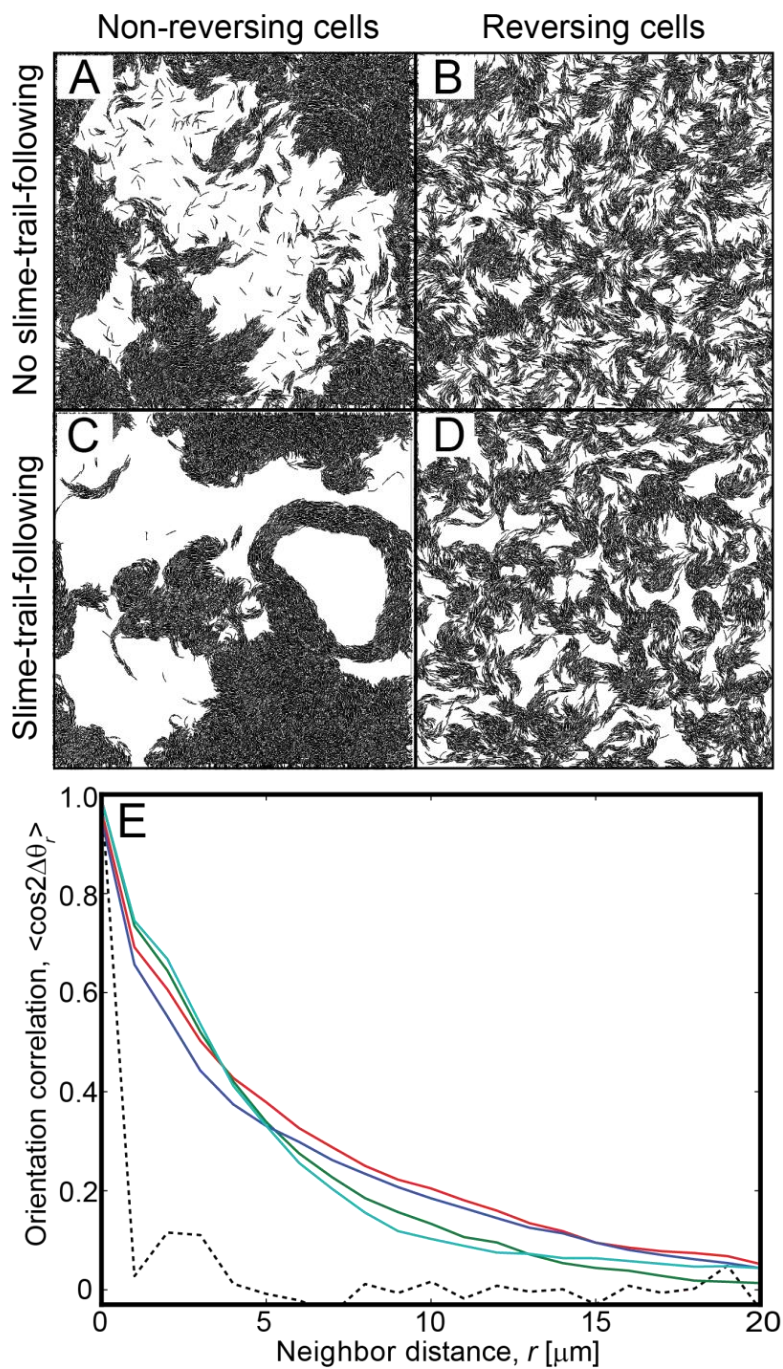

**Figure S6 Clustering behavior of cells at high cell densities** (A-D) Snapshots of simulation at 180 min for cell density $\eta = 0.60$. (E) Orientation correlation among cells at 180 min of simulation time for non-reversing cells (red), reversing cells (green), non-reversing cells with slime-trail-following (blue), and reversing cells with slime-trail-following (cyan). Dotted line represents the orientation correlation values at 1 min simulation time.

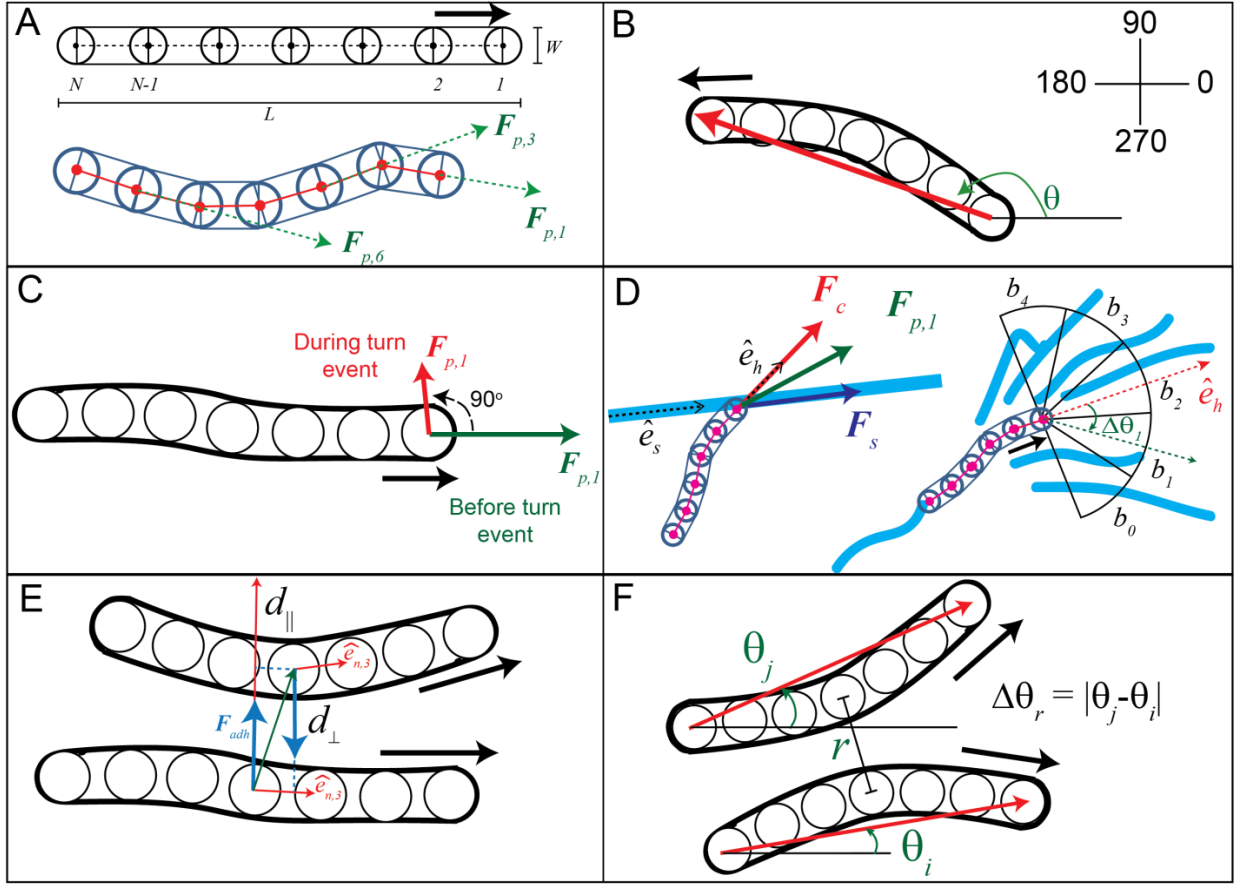

**Figure S7 Multi-segmented biophysical model of single *M. xanthus* cell as an agent in our simulation framework** (A) Each agent contains $N=7$ nodes connected by joints that simulate elastic behavior of the cell. Propulsive forces ($F_{p,i}$, green arrows) on the nodes, in the direction of next node, move the agent forward. (B) Orientation ($\theta$) of an agent defined as the angle made by the vector connecting from its tail node to head node with the X-axis. (C) Random noise in agent direction is introduced by reorienting the propulsive force ($F_{p,1}$) on its head node by $90°$ either clockwise or anti-clockwise randomly for a fixed amount of time (= 1 min). (D) Schematic for implementation of slime-trail following. When an agent encounters a slime trail, a part of the propulsive force on its head node ($F_s$) proportional to amount of slime in the trail is reoriented parallel to the direction of the slime-trail ($\hat{e}_s$). Remaining propulsive force $F_c$ ($=(F_T/(N-1)-|F_s|)\hat{e}_h$) acts in current head node direction ($\hat{e}_h$). Thus the resulting force on the head node $F_{p,1}$ maintains its magnitude but changes its direction due to its interaction with slime. In slime-rich regions (slime denoted by blue trails) of simulation, effective slime-trail direction ($\hat{e}_s$) is estimated by dividing a semi-circular slime search region at the head node of the agent into bins ($n=5$). $\hat{e}_s$ is chosen as the direction (center line) of the bin with high slime volume ($0.8 S_{max}$) but with least deviation ($\Delta\theta_s$) from current head node direction ($\hat{e}_h$). (E) Lateral adhesive forces ($F_{adh}$) between a pair of agents acting normal to node propulsion vectors ($\hat{e}_{n,i}$). These forces are implemented for simulations shown in Fig S5 only (F) Orientation correlation between a pair of agents, is computed by averaging $\cos(2\Delta\theta_r)$ over all agent pairs whose center nodes are separated by distance $r$. $\Delta\theta_r$ is the difference in orientations between the two agents.

Page 36 of 38

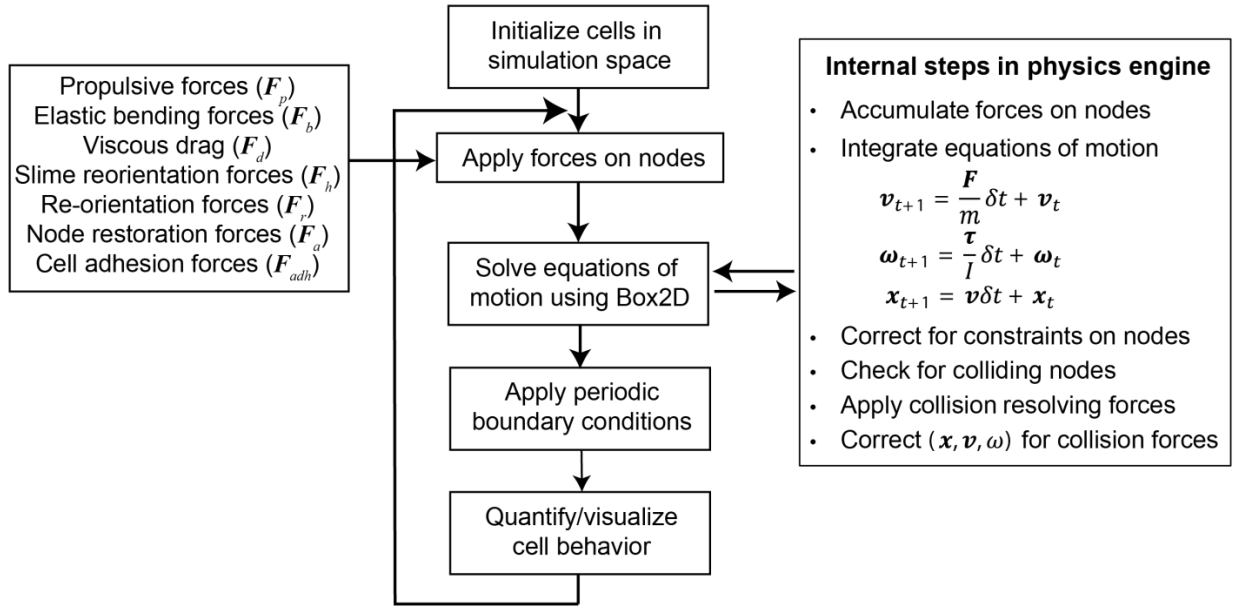

**Figure S8 Flow chart of simulation procedure for our agent-based-simulation framework**

**Movie S1. Evolution of clusters through agent collisions, merging and splitting of clusters**

**Movie S2. Clustering behavior of non-reversing agents in initial 60 min of simulation.** At the beginning, agents are initialized one by one over few time steps until desired cell density ($\eta = 0.24$) is reached. Units of time displayed here is min.

**Movie S3. Clustering behavior of periodically reversing agents in initial 60 min of simulation.** At the beginning, agents are initialized one by one over few time steps until desired cell density ($\eta = 0.24$) is reached. Units of time displayed here is min.

**Movie S4. Clustering behavior of periodically reversing agents following slime trails in initial 60 min of simulation.** Slime following mechanism parameters ($L_s = 11\,\mu m, \varepsilon_s = 1.0$). At the beginning, agents are initialized one by one over few time steps until desired cell density ($\eta = 0.24$) is reached. Units of time displayed here is min.

**Movie S5. Clustering behavior of non-reversing agents following slime trails after initial transition period of 60 min.** Slime following mechanism parameters ($L_s = 0.6\,\mu m, \varepsilon_s = 0.5$). Cell density $\eta = 0.24$. Units of time displayed here is min.

**Movie S6. Clustering behavior of periodically reversing agents following slime trails after initial transition period of 60 min.** Slime following mechanism parameters ($L_s = 11\,\mu m, \varepsilon_s = 0.2$). Cell density $\eta = 0.24$. Units of time displayed here is min.

**Movie S7. Circular cell aggregates formed by non-reversing agents with slime-following mechanism active.** 3% of all agents (represented as strings of nodes here) in the simulation are colored red to track individual agent movement inside the aggregate. Agents can slide past their neighbors inside the aggregate and move with approximately the same speed. As a result angular velocity of the agents near aggregate center is higher compared to agents farther from center.